# Propagation mechanism of surface-enhanced resonant Raman scattering light through one-dimensional plasmonic hotspot along silver nanowire dimer junction

Tamitake Itoh,[1]* Yuko S. Yamamoto[2], Jeyadevan Balachandran[3]

[1] Health and Medical Research Institute, National Institute of Advanced Industrial Science and Technology (AIST), Takamatsu, Kagawa 761-0395, Japan

[2] School of Materials Science, Japan Advanced Institute of Science and Technology (JAIST), Nomi, Ishikawa 923-1292, Japan

[3] Department of Engineering, the University of Shiga Prefecture, Hikone, Shiga 522-8533, Japan

*Corresponding author: tamitake-itou@aist.go.jp

## ABSTRACT

We investigate the propagation of surface-enhanced resonant Raman scattering (SERRS) light by several micrometers through a one-dimensional hotspot (1D HS) located between a plasmonic nanowire dimer (NWD). The propagation exhibits the properties, e.g. an effective propagation induced by excitation and detection polarization orthogonal to the 1D HS long axis, the propagation profiles composed of bright short and dark long propagations, SERRS spectral shapes independent of localized plasmon (LP) resonance of NWDs, redshifts in the SERRS spectra at the edges of 1D HSs, and considerable NWD-by-NWD variations in the propagation lengths. These properties are well reproduced by numerical calculations based on electromagnetism. These calculations reveals the following propagation mechanism: excitation light resonantly coupled with LP at the edges of 1D HSs, and the light energy is transferred to two types of junction SP modes supporting the short and long propagations; these modes are attributed to the upper and lower branches of coupled two SP modes. This mechanism comprehensively clarifies the abovementioned properties.

## I. INTRODUCTION

Effective cross-sections of light-matter interactions such as Raman scattering, fluorescence emissions, and their nonlinear counterparts are significantly enhanced or quenched by coupling with plasmons, which are collective excitations of conduction electrons of metallic nanostructures [1]. Plasmon-enhanced spectroscopies, which entail the enhancement of various molecular spectroscopies, exhibit the most significant enhancement at the gaps or junctions of plasmonic nanostructures because of the large electric field confined at these locations [1–4]. These locations are referred to as hotspots (HSs) of surface-enhanced (resonant) Raman scattering (SE(R)RS) spectroscopy, which is a plasmon-enhanced spectroscopy method that enables single-molecule Raman spectroscopy [5–7]. Various HSs have been developed using bright and dark plasmon modes, in addition to the localized plasmon (LP) and surface plasmon (SP) modes of nanostructures composed of nanoparticles (NPs), nanowires (NWs), nanosheets (NSs), and their complexes [8–11]. For example, HSs in an NP cluster generated by Fano resonance, which is the coupled resonance between bright and dark plasmon modes, significantly enhance and modify the SERRS spectra [8]. The HSs of NP dimer gaps enable single-molecule (SM)-SERRS detection by coupled LP modes [2,5,7]. The HSs at gaps or junctions between NWs and NPs facilitate the remote excitation of SERRS via coupling between LP and SP modes [9]. The HSs at the gaps between NPs or NWs on NSs serve as stable HSs owing to the mirror dipole effect [11]. Recently, HSs were proposed as locations for examining phenomena related to cavity quantum electrodynamics (QED), such as the strong coupling between molecular excitons and LPs, in addition to related phenomena such as exotic photochemical reactions [11–17].

The volumes of HSs at NP dimer gaps are limited to be only several cubic nanometers, thus resulting in unstable signals reflecting photo-induced thermal or chemical molecular fluctuations [18]. It was demonstrated that HSs at junctions of NW dimers are enlarged one-dimensionally by approximately 10,000 times (10 μm) [10]. The large volume of one-dimensional hotspots (1D HSs) is expected to resolve the problem of instability in the light–matter interactions. Furthermore, the 1D HSs may enable the remote excitation and propagation of SERRS light through extremely small gaps. The propagation of SP polaritons along the 1D gaps between two or more parallel NWs has been studied to clarify the gap SP modes supporting the propagation [19–26]. However, if the gap distance is less than 0.5 nm or the NWs are in contact [21–25], the gap SP modes supporting the propagation should be reconsidered, because the enhancement and propagation characteristics are largely different from those of the gap SP modes between NWs that are not in contact. Such cases are highly probable in experimentally prepared SERRS-active NWs [10]. This should be experimentally and theoretically investigated to understand the 1D HSs of SERRS better.

In this study, the propagation of SERRS light through 1D HSs located between silver nanowire dimers (NWDs) was investigated. The 1D HSs exhibit unique properties as follows. The SERRS light effectively propagates by excitation polarization orthogonal to the NWD long axis. Two types of channel-like propagation are observed along 1D HSs: one is the short and bright propagation, and another is the long and dark propagation. The SERRS spectra are rather independent of the LP resonance spectra of the NWDs, which is different from the case of NP dimers. The SERRS spectra at 1D HS edges exhibit redshifts from those at other positions. The propagation lengths largely vary from 1D HS to 1D HS. A comparison between these results and numerical calculations based on electromagnetism (EM) revealed that these properties originate from two types of junction SP modes. These SP modes correspond to the upper and lower branches of coupled SP modes between a dipole-dipole coupled mode and a quadrupole-quadrupole coupled mode by referring to the mode analysis of SPs of NWDs [23,24]. The upper and lower branch modes support the short and long channel-like propagation, respectively. The independence of LP resonance is due to excessive confinement of the enhanced electric field of SP modes within junctions; thus, the enhanced electric field is insensitive to the entire structure of the NWDs. The redshifts in SERRS spectra at the 1D HS edges are mainly attributed to the broader spectral shape of SP mode supporting the short propagation than that supporting the long propagation. The variations in propagation lengths are induced by the degree of mismatch between the SERRS spectra and the spectra of SP mode supporting the long propagation. These findings are helpful for optimizing enhancement and propagation through 1D HSs.

## II. MATERIALS AND METHODS

The silver NW samples were prepared as elucidated in [10]. The average NW diameter and length are 60 nm and 10 µm, respectively. The NW suspension was then dropped and dried on a glass plate. Thereafter, rhodamine 6G (R6G) methanol solution (~$1.0 \times 10^{-6}$ M) was dropped and dried on a glass plate. The effective concentration of the dye on the NWs was reduced by photobleaching most of the dye molecules adsorbed on the NW and glass surfaces using a green laser beam, which was confirmed by blinking signals of the SERRS and fluorescence from both the NWs and the glass surface [10]. The quality of the NWs was examined by the Fabry–Perot type resonance of longitudinal SP mode [27], thus indicating that the NWs can be treated as single-crystalline not as polycrystalline [10].

The dark-field images and two types of SERRS images were measured. The dark-field images were measured by focusing the white light of a 50-W halogen lamp through a dark-field condenser (NA 0.92) to selectively collect elastic scattering light [28]. The LP resonances of the NWDs were obtained from the spectra of elastic scattering light. The two types of SERRS images were measured to obtain images of the entire 1D HSs and 1D propagation images of SERRS light through the 1D HSs, as shown in Fig. 1(a). The SERRS images of the entire 1D HSs were measured by the wide-field circular polarized excitation of a green laser beam (532 nm, 3.5 W/cm$^2$, beam spot size 200 × 400 μm$^2$) loosely focused using an objective lens (N.A. 0.2). The SERRS images corresponding to 1D propagation were measured by the narrow-field linear polarized excitation of a green laser beam (532 nm, 50 W/cm$^2$, beam spot size 300 × 500 nm$^2$) tightly-focused using a 100× objective lens (LCPlanFl 100×, N.A. 0.9, Olympus, Tokyo) on the edge of the 1D HS. A 100× objective lens was used to collect both elastic scattering and SERRS light from the NWDs. The elastic scattering light and SERRS light from an identical NWD are sent to a polychromator for the spectroscopy through a pinhole to selectively measure the position which area is around 1.8×1.8 μm$^2$ on the glass surface. Figures S1(a)–S1(c) present the dark-field images, SERRS images obtained by wide-field excitation, and scanning electron microscopy (SEM) images, respectively. The SERRS activity of 30 NWDs was checked, and it was confirmed that all NWDs clearly exhibited 1D HSs. The isolated single NWs did not exhibit SERRS activity, thus indicating that the junctions of NWs play a critical role in 1D HSs. It was tentatively considered that the propagation is mainly induced by SERRS light and not by excitation light based on the synchronized blinking of SERRS light through 1D HSs (data not shown).

## III. RESULTS AND DISCUSSION

We observed the relationships between the dark-field, entire 1D HS, and 1D propagation images of SERRS light. First, we checked the 1D propagation of the excitation laser light using the isolated NWs. SPs cannot be directly coupled with light propagating free space because the momentum of a SP mode is always larger than that of light [29]. Thus, the excitation of SPs is carried out *via* excitation of LP at the edge of NW or emitters like fluorescent molecules on the NW, those of which can be directly coupled with the light [29]. Then the light energy is transfer from the LP mode (or emitters) to SP modes supporting the propagation through near-filed interaction. Figures 1(b1) and 1(b2) present the excitation light polarization dependence of 1D propagation. The excitation light polarized parallel to the NW long axis can resonate with the Fabry–Perot mode of the NW at the edge in Fig. S2(b) [27]. Then the light energy is provided to the SP mode. Thus, 1D propagation was observed for the excitation polarization parallel to the NW long axis as light emission from another edge. The SP modes supporting this propagation were extensively studied as monopole and higher-order modes [22,23,29]. In the case of excitation polarization orthogonal to the NW, the excitation light should resonate with the LP mode at the edge of the NW. However, the resonance energy of isolated NW is much higher than the excitation light energy ~2.3 eV (532 nm) as in Fig. S2(b), thus resulting in a lack of observation of 1D propagation. The propagation properties of SERRS light through 1D HSs are significantly different from those of isolated NWs. Figures 1(c1)–1(c4) and 1(d1)–1(d4) present the dark-field images and wide-field excitation SERRS images, respectively, for the four NWDs.

Several properties of 1D HSs are described below. The 1D HSs in Figs. 1(c4) and 1(d4) are observed along the NWD domain (yellow region in Fig. 1(c4) indicated by white arrow) and, the SERRS activity is not observed in the isolated NW domain (gray region in Fig. 1(c4) indicated by white arrow). Furthermore, two points of interest are observed in these images. First, the colors of the SERRS light are always green, although dark-field images exhibit various colors. Second, the edges of the SERRS light redshift from the centers of the 1D HSs. Figures 1(e1)–(e4) and 1(f1)–(f4) present the SERRS images with the narrow-field excitation polarized orthogonal and parallel to the 1D HSs, respectively. The locations of focusing light spots are the edges of 1D HSs. The SERRS light intensities around the excitation spots are much higher than those around the outside spots. This property is common for all 1D HSs, thus indicating that two types of SP modes are involved in the propagation of SERRS light. One supports bright and short-length propagation, and the other supports dark and long-length propagation. The propagating light intensities with orthogonal excitation in Figs. 1(e1)–(e4) are always higher than those with parallel excitation in Figs. 1(f1)–(f4). The images for both excitation polarizations exhibit similar propagation profiles. These properties are largely different from those of the propagation of isolated NWs in Figs. 1(b1) and 1(b2). These properties are discussed further using numerical calculations.

It was found that the polarization dependence of propagation of SERRS light along 1D HSs in Figs. 1(e1)–1(e4) and 1(f1)–1(f4) are opposite to those along the NW in Figs. 1(b1) and 1(b2). To further investigate the property, the polarization dependence of propagation was observed under Cross–Nicole conditions. Figures 2(a1)–2(a3) present a dark-field image and two entire 1D HS images measured by wide-field unpolarized excitation with detection polarizations orthogonal and parallel to the 1D HSs, respectively. The SERRS intensity becomes the maximum for the orthogonal excitation polarization. This property indicates that the propagation is induced by LP resonance orthogonal to the 1D HSs [10]. Figures 2(b1) and 2(b2) present the propagation for both excitation and detection polarizations orthogonal to the 1D HS, and for the excitation polarization orthogonal and detection polarization parallel to the 1D HS, respectively. Figure 2(b3) presents the detection polarization dependence of the SERRS intensity. The SERRS intensity shows the maximum for the orthogonal detection. Figures 2(c1) and 2(c2) present the propagation for the excitation polarization parallel and detection polarization orthogonal to the 1D HS, and for both excitation and detection polarizations parallel to the 1D HS, respectively. Figure 2(c3) presents the detection polarization dependence of the SERRS intensity. The SERRS intensity shows maximum for orthogonal excitation polarization. These identical polarization dependences irrespective of the excitation polarization in Fig. 2(b3) and 2(c3) indicate that the propagation is supported by SP modes orthogonally polarized to the 1D HSs. Thus, in the case of parallel excitation, the propagation is occurred by the orthogonally polarized SP modes coupled with the parallel LP modes via. near-field interaction at the edges. These polarization properties are discussed further using numerical calculations.

The SERRS images of 1D HSs in Figs. 1(d1)–1(d4) always exhibit monotonous green-like colors, although their dark-field images in Figs. 1 (c1)–1(c4) exhibit color variations. The differences in the NWD-by-NWD variations in colors between the SERRS and dark-field images were quantitatively investigated using the LP resonance and SERRS spectroscopy, respectively. Figures 3(a1)–3(a3) and 3(b1)–3(b3) present the typical LP resonance and SERRS spectra, respectively, for the three NWDs. The SERRS spectra are accompanied by large background emission, which is attributed to surface-enhanced fluorescence (SEF) of R6G molecules [30]. Here, SERRS with SEF is simply referred to SERRS. The LP resonance and SERRS spectra were obtained from the center positions of the 1D HSs by wide-field excitation. The LP resonance spectra clearly exhibit NWD-by-NWD variations in the peak wavelengths. However, the SERRS spectra did not exhibit such variations. The lack of variations is inconsistent with the electromagnetic (EM) mechanism of SERRS, in which SERRS light is emitted through LP resonance [4]. The EM mechanism approximately describes the SERRS cross-section spectrum $\sigma_{\text{SERRS}}$ as follows:

$$\sigma_{\text{SERRS}}(\lambda_{\text{ex}}, \lambda_{\text{em}}) = \left| \frac{E^{\text{loc}}(\lambda_{\text{ex}})}{E^{\text{I}}(\lambda_{\text{ex}})} \right|^2 \left[ \sigma_{\text{RRS}}(\lambda_{\text{ex}}, \lambda_{\text{em}}) + q\sigma_{\text{FL}}(\lambda_{\text{ex}}, \lambda_{\text{em}}) \right] \left| \frac{E^{\text{loc}}(\lambda_{\text{em}})}{E^{\text{I}}(\lambda_{\text{em}})} \right|^2, \,(1)$$

where $\lambda_{\text{ex}}$ and $\lambda_{\text{em}}$ are the excitation and emission wavelengths, respectively; $E^{\text{I}}$ and $E^{\text{loc}}$ are the amplitudes of a far-field and a LP-enhanced local field, respectively; and $\sigma_{\text{RRS}}$, $q$, and $\sigma_{\text{FL}}$ are the resonance Raman scattering cross-section, fluorescence quenching factor, and fluorescence cross-section spectrum of a molecule free from enhancement, respectively [30]. The EM mechanism was verified using single silver NP dimers, which allowed for the direct observation of the relationships between LP resonance and SERRS. The relationships explained the polarization dependence, $Q$ factor dependence, $\lambda_{\text{ex}}$ dependence, and $\lambda_{\text{em}}$ dependence of SERRS based on the EM mechanism [31]. The spectral shape of $E^{\text{loc}}$ is approximately identical to the LP resonance exhibited in the elastic scattering spectra of silver NP dimers [32]. Thus, the term $\left| E^{\text{loc}}(\lambda_{\text{em}}) / E^{\text{I}}(\lambda_{\text{em}}) \right|^2$ in Eq. (1) indicates that the SERRS spectral shape is modulated by the LP resonance. In particular, a clear relationship between SERRS spectral envelopes and LP resonance spectra was reported for silver NP dimers [30,32]. To confirm this inconsistency against the previous reports, the relationship was checked for 20 NWDs as shown in Fig. 3(c). The peak wavelengths of LP resonance ($\lambda_{\text{LP}}$) are distributed from 530 to 610 nm; however, the peak wavelengths of the SERRS spectral envelopes ($\lambda_{\text{SERRS}}$) are distributed only from 545 to 565 nm. This independence of the SERRS spectra on LP resonance should be clarified based on the EM mechanism.

The color variations within SERRS 1D HSs are shown in Figs. 1 (d1)–1(d4) and 1(e1)–1(e4). To quantitatively analyze the variation, the position dependence of the SERRS spectra of single 1D HSs was investigated. As indicated in Figs. 4(a1) and 4(a2), the colors of the edges of the 1D HSs are always redshifted from their centers. The redshifts were commonly observed in both wide- and narrow-field excitation SERRS images. Thus, the SERRS spectra of the edges and centers were investigated by wide-field excitation. Figure 4(b) presents the SERRS spectra from the edge (red color) and center (green color). Both spectra are normalized at peak intensities. The relative SERRS intensity at around 600 nm of the edges is higher than that of the centers as shown in Fig. 4(b). This trend was verified using 20 NWDs. Figure 4(c) presents the ratios of the SERRS intensities at 550 nm to those at 600 nm at the edges with respect to the ratios at the centers. It can be noted that the ratios at the edges are always smaller than those at the centers, thus indicating that this trend of redshifts is common for the SERRS spectra. The trend suggests that the spectrum of SP mode supporting the propagation around the edges is different from that at the center. Thus, the position-dependent spectral shapes are discussed further using numerical calculations with SP mode analysis based on Eq. (1).

Figures 1(e1)–1(e4) present the NWD-by-NWD variations in the propagation profiles of SERRS light through 1D HSs. As shown in Fig. 1, the propagation profiles are likely composed of short propagation < ~1.0 μm with high intensity and long propagation > ~2.0 μm with low intensity, thus indicating that two types of SP modes support the propagation. To quantitatively analyze the variations, the propagation profiles were evaluated. Figures 5(a1)–5(a3) present the propagation profiles with the SERRS images for the three NWDs by wide-field excitation

(dotted lines) and narrow-field excitation (solid lines). The wide- and narrow-field excitation SERRS images reveal the entire lengths of the 1D HSs and the propagation lengths of the SERRS light along the 1D HSs, respectively. The propagation length ($L_p$) is tentatively defined as the length at which the SERRS intensity is 3% of the maximum intensity at the excitation spot. Figure 5(a1) reveals that SERRS light propagates by 5 μm. However, Figs. 5(a2) and 5(a3) reveal that SERRS light propagates by only ~2.5 μm and ~1.5 μm, respectively. Figure 5(b) presents the relationship between $L_p$s and the entire lengths of the 1D HSs. The values of $L_p$ is limited by the loss due to scattering and absorption by factors such as the surface roughness, domain boundaries, and defects [26]. Thus, the variations in $L_p$s may be due to the degree of these factors. However, other factors that influence the $L_p$s should be investigated to better understand the variations in $L_p$s with respect to the importance of the length for the applications.

To clarify the abovementioned properties of the propagation of SERRS light along 1D HSs, as presented in Figs. 1–5, the experimental results were analyzed by classical electromagnetism using a finite difference time domain (FDTD) method. The experimentally obtained properties are summarized below.

(1) The SERRS light effectively propagates by excitation polarization orthogonal to 1D HSs.

(2) The propagating SERRS light is mainly polarized orthogonal to 1D HSs, inclusive of the case with excitation polarization parallel to 1D HSs.

(3) Two types of propagation were observed along 1D HSs: short propagation < ~1.0 μm with high intensity and long propagation > ~2.0 μm with low intensity.

(4) The SERRS spectra are independent of the LP resonance spectra of the NWDs.

(5) The SERRS spectra of the 1D HS edge positions redshift from the SERRS spectra of the center positions.

(6) The propagation lengths considerably vary 1D HS to 1D HS.

The abovementioned six-fold properties were investigated using the FDTD calculation. Figure 6(a) presents the calculation conditions for reproducing the experimental conditions in Fig. 1(a). The dielectric function of silver was obtained from [33]. The beam diameters of the wide-field and narrow-field excitations are infinite and 400 nm, respectively. The excitation position of the narrow-field excitation is set at the edge of the 1D HS. The lengths of the two NWs are 5.5 and 4.5 µm, respectively. Thus, the length of the 1D HS, which is a gap or junction, is 4.5 µm. The diameter ($D_{NW}$) of the NW and the gap distance ($d_g$) between the two NWs are parameters used for examining 1D propagation. The amplitude of the incident electric field $\left| E^1 \right|$ is set as 1 V/m. Thus, the value of the calculated amplitude $\left| E^{loc} \right|$ is the same as the enhancement factor (EF) of electric field amplitude as $\left| E^{loc} / E^1 \right|$ in Eq. (1). Figure 6(b) presents the calculated scattering cross-section ($\sigma_{sca}$) spectra of wide-field excitation with excitation light polarized parallel and orthogonal to the NWD long axis. Typical values $D_{NW} = 60$ nm and $d_g = 0$ nm are used for the calculation. The LP resonance peaks are observed at 550 nm and 425 nm for the orthogonal excitation polarization. These peaks correspond to the dipole and quadrupole LP modes, respectively. No plasmon resonance peak is observed for the parallel excitation polarization. This polarization dependence is consistent with the results in Fig. 3 and those of a previous report [10]; this indicates that the LP resonance at 550 nm is the origin of the enhancement of the electric field propagating along the 1D HS. Thereafter, the propagation of excitation light at $\lambda_{ex} = 550$ nm was investigated by narrow-field excitation. Figures 6(c1)–6(e1) present the setups for calculation for the electric field distributions in the x-y ($z = 2.5$ µm), y-z ($y = -9$ nm), and x-z ($x = 0$ nm) planes. The position $z = 0$ is the starting point of 1D propagation.

The $x$ and $y$ positions are determined to show the largest EF values. We consider that the propagation is mainly induced by SERRS light generated at the 1D HS edge because of the synchronized blinking of SERRS light through 1D HSs (data not shown). Thus, we assumed that $\lambda_{ex}$ is treated as $\lambda_{em}$ in the numerical analysis. Figures 6(c2)–6(e2) present the total electric field $E_{all}^{loc}\left(x, y, z, \lambda_{ex}\right)=\sqrt{E_x^{loc2}+E_y^{loc2}+E_z^{loc2}}$ ($E_{all}^{loc}=E^{loc}$ in Eq. (1)) distributions with the excitation polarization orthogonal to the NWD long axis. Figures 6(c3)–6(e3) present $E_{all}^{loc}$ distributions with excitation polarization parallel to the NWD long axis. The red regions indicate that $E_{all}^{loc}$ exhibited EFs $\geq 1$. The calculations in Figs. 6(c2)–6(e2) clearly reveal that the propagation of enhanced $E_{all}^{loc}$ only occurs along the 1D HS with the orthogonal excitation. The propagation of weak $E_{all}^{loc}$ occurs even with the parallel excitation in Figs. 6(c3)–6(e3). The EF for the orthogonal excitation is smaller than that for the parallel excitation by ~5 times at around $z = 2.5$ μm. These calculations are consistent with the experimentally obtained polarization properties in Figs.1(e1)–1(e4) and 1(f1)–1(f4), thus indicating that the calculations reproduce the experimentally observed propagation of electric fields along the 1D HS. The identical polarization dependence between the LP resonance in Fig. 6(b) and the propagating enhanced electric field indicate that the LP resonance at 550 nm is the origin of the propagation of SERRS light. The orthogonal polarization of the propagating $E_{all}^{loc}$ indicates that the oscillating charges of SP mode on the two NWs are out of phase. Thus, the SP modes supporting the propagation may be similar to the monopole-monopole (MM) coupled or the dipole-dipole (DD) coupled modes of NWDs [23-26]. Thus, two propagation mechanisms are considered. The first is for the orthogonal excitation: the excitation light is coupled with the orthogonally polarized LP resonance at the edge and then the light energy is efficiently transferred to the SP modes supporting the propagation. The second is for parallel excitation polarization: the excitation light is first coupled with the parallelly polarized LP mode at the edge and then the light energy is inefficiently transferred to the SP modes supporting the propagation *via* near-field interaction. By the way, the fluctuating electric field pattern with the period of ~2 μm around the NWD in Figs. 6(d2) and 6(d3) suggested that chiral SP propagation is induced by the asymmetric edge structure of the 1D HS [34]. Indeed, such electric field pattern disappears for NWDs with the asymmetric edge structure (data not shown). The chiral propagation along a 1D HS is intriguing. But we do not discuss it in this study.

As shown in Fig. 2, the polarization dependence of propagation of SERRS light along 1D HSs is identical irrespective of excitation polarization directions under Cross–Nicole conditions. To understand the result, the polarization dependence of the propagating enhanced electric fields was calculated under Cross–Nicole conditions. Figures 7(a1)–7(a3) present $E_{all}^{loc}$ , $E_{x}^{loc}$, and $E_{z}^{loc}$, respectively, at $x = 0$ nm and $y = $ -9 nm for the excitation polarization orthogonal to the NWD long axis at $\lambda_{ex} = 550$ nm. The positions $x$ and $y$ are selected to exhibit the highest EFs in the x-y plane. The calculation indicated that $E_{x}^{loc} >> E_{z}^{loc} >> E_{y}^{loc}$. Thus, $E_{all}^{loc}$ is mainly determined by $E_{x}^{loc}$. The profile in Fig. 7(a1) indicates the two types of propagation. The first is the large EF of ~30 with a short propagation length of ~1.0 μm (red frame). The second is the small EF of ~5 with a long propagation length of ~1.0 to 4.5 μm (blue frame). These types of propagation are consistent with the experimental results in Fig. 1. Figures 7(b1)–7(b3) present the $\lambda_{ex}$ dependence of the propagation profiles of $E_{all}^{loc}$, $E_{x}^{loc}$, and $E_{z}^{loc}$, respectively, expressed as contour maps. It should be noted that the enhanced $E_{x}^{loc} \geq 1$ for long propagation is observed only at $\lambda_{ex}$ ~550 nm for $z > 1$ μm, thus indicating that the SP mode supporting the long propagation has a spectral maximum at ~550 nm. This restriction of $\lambda_{ex}$ ~550 nm is consistent with the experimental results that SERRS spectra always exhibit $\lambda_{SERRS}$ at ~550 nm in Figs. 3(b1)–3(b3). The value of 550 nm is also identical to the calculated $\lambda_{LP}$ in Fig. 6(b). Thus, the light energy coupled with LP resonance can efficiently transfer to the SP modes supporting the long propagation. The $\lambda_{ex}$ dependence of the propagation profile of $E_z^{loc}$ in Fig. 7(b3) is similar to that of an isolated NW in Fig. S2(f1), thus indicating the superposition of the uncoupled monopole (M) and dipole (D) modes. The EFs of SP modes supporting the short and long propagation in Fig. 7(a1) are ~30 and ~5, respectively. The two EFs correspond to the SERRS enhancement factors of $8.1 \times 10^5$ and $6.3 \times 10^2$ regarding $\left| E^{loc} / E^1 \right|^4$ in Eq. (1), respectively. This large difference in EFs between the two SP modes indicates that the SERRS light field energy propagating 1D HSs are mainly generated by the SP mode supporting the short propagation, not by the SP mode supporting the long propagation.

Thereafter, the excitation polarization was rotated to parallel to the NWD long axis to check the results of Cross–Nicole conditions in Fig. 2(c1)–2(c3). Figures 7(c1)–7(c3) present $E_{all}^{loc}$, $E_x^{loc}$, and $E_z^{loc}$ at $y$ = -9 nm, respectively, for $\lambda_{ex}$ = 550 nm. As can be seen from Figs. 7(c2) and 7(c3), $E_x^{loc}$ is larger than $E_z^{loc}$, thus indicating that $E_{all}^{loc}$ is mainly determined by $E_x^{loc}$ even the parallel excitation polarization. Figures 7(d1)–7(d3) present the $\lambda_{ex}$ dependence of the propagation profiles of $E_{all}^{loc}$, $E_x^{loc}$, and $E_z^{loc}$, respectively, expressed as contour maps. The $\lambda_{ex}$ dependence supports the indication that the propagation profile of $E_{all}^{loc}$ is mainly determined by $E_x^{loc}$. This result is consistent with the experimental results in Figs. 2(c1)–2(c3). The LP resonance having $\lambda_{LP}$ = 550 nm in Figs. 6(b) cannot be directly excited by the parallel polarization. Thus, the propagation mechanism is considered as follows: the excitation light is first coupled to the LP resonance which is parallelly polarized to NWD long axis at the edge, and the light energy is then transferred to the orthogonally polarized SP modes supporting the propagation *via* near-field interaction. The LP resonance appears as a flat spectrum in Fig. 6(b). Thus, the coupling efficiency between the excitation light and the LP modes under the parallel excitation polarization is much inefficient than that under orthogonal excitation polarization. The $\lambda_{ex}$ dependence of $E_z^{loc}$ in Fig. 7(d3) is similar to that of $E_z^{loc}$ of the monomer NW in Fig. S2(f2), thus indicating the superposition of uncoupled M and D modes.

We considered in the discussion of Fig. 6 that the SP modes supporting the propagation of the SERRS light along 1D HSs have similar properties to the MM- and DD-coupled modes. The concept of MM- and DD-coupled modes supposes two NWs separating from each other as $d_g > 0$ nm [23]. However, it is highly probable that the NWDs showing strong SERRS signals are in contact as $d_g = 0$ nm with referring to the SEM images in Figs. S1(c1)–S1(c4). Thus, the applicability of the concept to the NWD contacting each other should be examined by changing the gap distances. Thus, the propagation profiles between the two NWs were evaluated by changing $d_g$ from 20 to –5 nm. It should be noted that the region $d_g = 0$ to 1 nm was excluded from the examination to avoid the effect of charge transfer SPs between the gaps [24,25]. Figures 8(a1)–8(a5) show the propagation profiles of $E_{all}^{loc}(\lambda_{ex})$ for $\lambda_{ex}$ = 400 to 800 nm expressed as contour maps by changing $d_g$ from 20 to -5 nm. The excitation polarization is orthogonal to the NWD long axis. Figures 8(b1)–8(b5) show the $\sigma_{sca}$ spectra by wide-field excitation (black curves) and the spectra of $E_{all}^{loc}$ at $x = 0$ and $z = 2.5$ µm (red curves), respectively. The values of $y$ are selected to exhibit the maximum EFs in the x-y plane. The wavelengths at maximum $E_{all}^{loc}(\lambda_{ex})$, as shown in Figs. 8(b3)–8(b5), are referred to as $\lambda_{MAXS}$. Figures 8(c1a)–8(c5a) and 8(c1b)–8(c5b) present the cross-sections of $\left| E_{\text{all}}^{\text{loc}} / E_{\text{max}}^{\text{loc}} \right|$ for the x-y plane at $z = 100$ nm (red frames) and 2.5 μm (black frames), respectively, where $E_{\text{max}}^{\text{loc}}$ indicates the maximum value of $E_{\text{all}}^{\text{loc}}$ in the x-y plane. We discuss SP modes supporting the short and long propagation based on the calculations in Fig. 8.

The propagation profiles in Figs. 8(a1)–8(a5) changed dramatically by crossing $d_{\text{g}} = 0$ nm. The propagation profiles for $d_{\text{g}} \geq 1$ nm exhibit the oscillation structures with the periods dependent on $\lambda_{\text{ex}}$ in Figs. 8(a1)–8(a2). The oscillation periods and decay lengths for $d_{\text{g}} \geq 1$ nm are consistent with the MM- and DD-coupled modes [23]. It was clearly observed in Figs. 8(a1) and 8(a2) that the oscillation periods and decay lengths are smaller and shorter with decreasing $d_{\text{g}}$, respectively, as summarized in Fig. 8(d). These changes in Fig. 8(d) are induced by the confinement of mode volumes by decreasing $d_{\text{g}}$ [23]. Indeed, the number of oscillations until the amplitude was $1/e$ is commonly eight for all $d_{\text{g}}$s $\geq 1$ nm. Such oscillation structures are compressed around $z = 0$ nm as $d_{\text{g}}$ decreased in Fig. 8(d) and almost disappear at $d_{\text{g}} = 0$ nm in Figs. 8(a3). For $d_{\text{g}} \leq 0$ nm, channel-type propagation appears at a specific $\lambda_{\text{ex}}$ in Figs. 8(a3)–8(a5) instead of the oscillation structures. Thus, channel-type propagation cannot be supported by the MM- and DD-coupled modes. The analysis of the SP modes of NWDs for $d_{\text{g}} \leq 0$ nm has been carried out by several groups [19,21,24,25]. However, propagation profiles supported by these SP modes have not been provided.

Thus, we discuss the SP modes supporting the channel-type propagation in Figs. 8(a3)–8(a5) based on [23, 24]. The channel-type propagation is composed of short and long propagation as shown in Fig. 7(a1). We use the eigenmode analysis of gap SP modes in [23] and the dispersion relations of gap and junction SP modes in [24] for the discussion. The dispersion relation of MM-coupled modes in [24] shows that the momentum of wave vector parallel to NWD long axis becomes too large to exhibit effective propagation around $d_g$ ~0 nm (Figs. S3(a1)–(a4) and Fig. S3(b)). This result is consistent with our calculations showing the compression of oscillating structures for $d_g \leq 2$ nm in Fig. 8(d) and Fig. S3(b). Thus, the SP mode supporting short propagation mode in our calculation is not MM-coupled modes. The dispersion relation of SP modes in [24] indicates that the DD-coupled mode starts to interact with a quadrupole and quadrupole (QQ)-coupled mode as decreasing $d_g$ ~ 0 nm (Figs. S3(a1)–S3(a4)). As the result of interaction, mode splitting occurs at the crossing point between DD- and QQ-coupled modes at the photon energy region ~2.4 eV (~530 nm) for $d_g$ = -1 nm [24] (Figs. S3(a2) and S3(a3)). The upper and lower branch of the coupled SP mode appear as two peaks around 530 nm in the spectrum the local density of optical states (LDOS) in [24]. The positions of the two peaks are approximately consistent with $\lambda_{MAX}$ ~525 nm for $d_g$ = -1 nm in Fig. 8(f). The lack of two peaks in spectra of $E_{all}^{loc}(\lambda_{ex})$ in Figs. 8(b3)–8(b5) may be due to the low spectral resolution 25 nm in our calculation. This consistency among [24], the calculation, and the experiments indicates that the short and long channel-type propagation in Figs. 8(a3)–8(a5) are supported by the upper and lower branch modes of the coupled SP mode between the DD- and QQ-coupled mode, respectively.

We discuss changes in $\lambda_{LP}$ and $\lambda_{MAX}$ in Figs. 8(a3)–8(a5). For $d_g \geq 1$ nm, $\lambda_{LP}$s in the $\sigma_{sca}$ spectra in Fig. 8(b1) and 8(b2) are shorter than 400 nm, and the spectra of $E_{all}^{loc}(\lambda_{ex})$ exhibit complex structures reflecting the oscillation structures as in Figs. 8(a1) and 8(a2). For $d_g \leq 0$ nm in Fig. 8(b3)–8(b5), $\lambda_{LP}$s are longer than 400 nm and exhibit blueshifts with an increase in the overlapping of the two NWs. Blueshifts in $\lambda_{MAX}$ are also observed with $\lambda_{LP}$ in Fig. 8(b3)–8(b5). The relationship between $\lambda_{LP}$s and $d_g$s from 0 to -10 nm is summarized in Fig. 8(e). The relationship between $\lambda_{MAX}$s and $d_g$s from 0 to -10 nm is summarized in Fig. 8(f). The common blueshift in $\lambda_{LP}$ and $\lambda_{MAX}$ can be confirmed with respect to a decrease in $d_g$, thus suggesting that the light energy resonating with the LP mode with $\lambda_{LP}$ is efficiently injected into the SP mode with $\lambda_{MAX}$ supporting the channel-type propagation in Figs. 8(a3)–8(a5).

The properties of SP modes (e.g., monopole or dipole) clearly appear in the cross-sections of $\left| E_{all}^{loc} / E_{max}^{loc} \right|$ as in Figs. S2(c2a) and S2(c2b). Thus, the cross-sections of $\left| E_{all}^{loc} / E_{max}^{loc} \right|$ clarify the properties of SP modes supporting the propagation along 1D HSs. Figures 8(c1)–8(c5) show that the cross-sections changed significantly with a decrease in $d_g$. The $\left| E_{all}^{loc} / E_{max}^{loc} \right|$ in the cross-sections of the NWDs are consistent with the theoretical analyses conducted by several groups [19,21,23,24]. The cross-sections at $z = 100$ nm in Fig. 8(c1a) and 8(c2a) for $d_g \geq 1$ nm exhibit large values of $\left| E_{all}^{loc} / E_{max}^{loc} \right|$ inside the NWs. This distribution is the property of monopole mode [23], thus indicating that these cross-sections mainly exhibit the MM-coupled mode. Such distribution of $\left| E_{all}^{loc} / E_{max}^{loc} \right|$ inside NWs disappears within $z \sim 1$ μm due to the large ohmic loss of the metal. Alternatively, the different distribution appears at $z = 2.5$ μm. The cross-sections at $z = 2.5$ μm in Figs. 8(c1b) and 8(c2b) for $d_g \geq 1$ nm do not exhibit large values of $\left| E_{all}^{loc} / E_{max}^{loc} \right|$ inside the NWs. Instead, $\left| E_{all}^{loc} / E_{max}^{loc} \right|$ appears on both the sides and centers of the NWDs. These distributions are the properties of DD-coupled mode [23]. The propagation length of DD-coupled mode is much larger than that of MM-coupled mode because of the smaller ohmic loss. These results are consistent with the previous reports [23]. Thus, these two types of coupled SP modes support the propagation in the case of $d_g \geq 1$ nm.

However, for $d_g = $ -2 to 0 nm in Figs. 8(c3)–8(c5), the $\left| E_{all}^{loc} / E_{max}^{loc} \right|$ in the cross-sections dramatically changes. We attributed the SP modes supporting the short and long channel-type

propagation for $d_g \leq 0$ nm to the upper and lower branch of the coupled SP modes between the DD- and QQ-coupled mode, respectively, based on [23,24] with Fig. S3. Thus, we investigated the properties of these coupled SP modes. The $\left| E_{all}^{loc} / E_{max}^{loc} \right|$ in the cross-sections are tightly localized around the junctions as shown in Figs. 8(c3) and 8(c4). This property resembles the channel SP mode of the 1D tapered V groove on a metal surface [35]. Thus, the blueshifts in $\lambda_{LP}$ and $\lambda_{MAX}$ with respect to the decrease in $d_g$ from 0 to -10 nm in Figs. 8(e) and 8(f) can be explained by increasing the taper angle of the junction by overlapping two NWs [35]. The insets in Figs. 8(c3a) and 8(c3b) present enlarged cross-sections around the NWD junctions. The $\left| E_{all}^{loc} / E_{max}^{loc} \right|$ in Fig. 8(c3a) is more distributed inside the NWs than that in Fig. 8(c3b). The difference in distributions explains that the propagation length of the lower branch of the coupled SP mode is longer than the upper branch of the coupled SP mode. For $d_g \leq -5$ nm in Fig. 8(c5), the distribution of $\left| E_{all}^{loc} / E_{max}^{loc} \right|$ change again. The $\left| E_{all}^{loc} / E_{max}^{loc} \right|$ at $z = 100$ nm is observed inside the NWs as being similar to the M mode in Fig. S2(c2a). Thus, the SP mode supporting the short propagation retains a feature of the M mode. This distribution disappears within $z \sim 1$ μm owing to the large ohmic loss. By contrast, the cross-sections at $z = 2.5$ μm in Fig. 8(c5b) do not exhibit $\left| E_{all}^{loc} / E_{max}^{loc} \right|$ inside the NWs, and $\left| E_{all}^{loc} / E_{max}^{loc} \right|$ is observed on both sides and the junction of the NWDs. These distributions are similar to the property of the D mode in Fig. S2(c2b) [23]. The appearance of the properties of M and D modes in $\left| E_{all}^{loc} / E_{max}^{loc} \right|$ by decreasing $d_g$ is reasonable, because the SP modes supporting the short and long propagation become M and D modes for $d_g$ = -60 nm.

We summarize the discussion in Fig. 8 as follows. The SP modes supporting the long and short propagations changed from the MM- and DD-coupled modes, respectively, for $d_g \geq 1$ nm in Fig. 8(c1) and 8(c2) to the upper and lower branch of coupled SP modes between DD- and QQ-coupled mode, respectively, for -2 ≤ $d_g$ ≤ 0 nm in Fig. 8(c3) and 8(c4) to the M and D modes, respectively, for $d_g$ ≤ -5 nm in Fig. 8(c5). The trends are acceptable because the cross-sections should exhibit the M or D mode for $d_g = \infty$ or -60 nm. Finally, the calculation results are compared with the experimental results in Fig. 3. The experimentally obtained $\lambda_{LPS}$ and $\lambda_{SERRS}$ are plotted as dotted lines in Fig. 8(e) and Fig. 8(f), respectively. The experimental results are approximately consistent with the calculations by assuming $D_{NW}$ ~60 nm and $d_g$ ~0 nm. Furthermore, the lack of clear oscillation structures in experimentally obtained propagation profiles in Fig. 5 agrees with the calculated channel-type propagation profiles. Thus, the SP modes supporting the experimentally observed short and long propagations are mainly attributed to the upper ($z <$ ~1 μm) and lower ($z >$ ~1 μm) branch of the coupled SP modes between the DD- and QQ- coupled mode, respectively. The dispersion relations of coupled SP modes in [24] are quantitatively consistent with our attribution.

We discuss the NW diameter dependence of the SP modes supporting the propagation of SERRS light along 1D HSs. In Fig. 8, the experimental results are approximately reproduced by assuming $D_{NW}$ ~60 nm and $d_g$ ~0 nm. However, the experimental results in Fig. 3 revealed that $\lambda_{LP}$ exhibited large variations, although $\lambda_{SERRS}$ exhibited minor variations. This difference in the variations is inconsistent with SERRS EM mechanism, in which $\lambda_{LP}$ is approximately the same as $\lambda_{SERRS}$ because the term $\left| E^{loc}(\lambda_{em}) / E^{1}(\lambda_{em}) \right|^2$ ($= \left| E^{loc}_{all}(\lambda_{em}) / E^{loc}_{max}(\lambda_{em}) \right|^2$) in Eq. (1) indicates that the SERRS spectrum is modulated by LP resonance [30-32]. To clarify this inconsistency, the propagation profiles of the enhanced electric field along 1D HSs are examined by changing $D_{NW}$ from 20 to 140 nm with $d_g = 0$ nm. The excitation polarization for all the calculations in Fig. 9 is orthogonal to the NWD long axis. Figures 9(a1)–9(a5) present the $\lambda_{ex}$ dependences of propagation profiles of $E_{\text{all}}^{\text{loc}}\left(\lambda_{\text{ex}}\right)$ expressed as contour maps. All profiles exhibit channel-type propagation at specific $\lambda_{\text{MAX}}$s. The $\lambda_{\text{MAX}}$ in the channel-type propagation of $E_{\text{all}}^{\text{loc}}\left(\lambda_{\text{ex}}\right)$ at $z = 2.5$ μm redshift from 450 to 550 nm with an increase in $D_{\text{NW}}$ from 20 to 60 nm as in Figs. 9(a1)–8(a3). It should be noted that the $\lambda_{\text{MAX}}$ does not increase when $D_{\text{NW}}$ is increased from 60 to 100 nm, as in Figs. 9(a3)–8(a6). This trend of $\lambda_{\text{MAX}}$ may resolve the inconsistency in the variations in $\lambda_{\text{LP}}$ and $\lambda_{\text{SERRS}}$ in Fig. 3. Thus, these properties are checked using the spectra of $\sigma_{\text{sca}}$ and $E_{\text{all}}$. Figures 9(b1)–9(b5) present the spectra of $\sigma_{\text{sca}}(\lambda_{\text{ex}})$ (black curves) by wide-field excitation and the spectra of $E_{\text{all}}^{\text{loc}}\left(\lambda_{\text{ex}}\right)$ (red curves) at $z = 2.5$ μm. For the $D_{\text{NW}} = 20$ to 60 nm, both $\lambda_{\text{LP}}$ and $\lambda_{\text{MAX}}$ commonly redshift from 450 to 550 nm. For the $D_{\text{NW}} = 60$ to 100 nm, $\lambda_{\text{LP}}$ exhibit redshift from 550 to 650 nm; however, $\lambda_{\text{MAX}}$ is maintained at 550 nm. The range of $\lambda_{\text{MAX}}$ for $D_{\text{NW}} = 60$ to 100 nm can explain the different trends between $\lambda_{\text{LP}}$ (redshifting) and $\lambda_{\text{MAX}}$ (maintained).

We discuss the mechanism of the different trends between $\lambda_{\text{LP}}$ and $\lambda_{\text{MAX}}$. We compared these trends with the distribution of electric fields on the x-y plane of NWDs. Figures 9(c1a)–9(c5a) and 9(c1b)–9(c5b) present the cross-sections of $\left|E_{\text{all}}^{\text{loc}} / E_{\text{max}}^{\text{loc}}\right|$ on the x-y plane at $z = 100$ nm and 2.5 μm, respectively, for $D_{\text{NW}} = 20$ to 100 nm. For $D_{\text{NW}} < 60$ nm as in Figs 9(c1b)–9(c3b) $\left|E_{\text{all}}^{\text{loc}} / E_{\text{max}}^{\text{loc}}\right|$ is distributed around both sides of the NWDs. This distribution whole around the NWDs indicates that the spectra of SP modes supporting the propagation are determined by the entire structures of the NWDs. However, for $D_{\text{NW}} > 60$ nm as in Figs. 9(c4b)–9(c5b) this distribution around both sides of the NWDs almost disappears and $\left|E_{\text{all}}^{\text{loc}} / E_{\text{max}}^{\text{loc}}\right|$ is localized only around junctions. The disappearance and localization indicate that the electric field propagating through the 1D HS "feels" only the structure inside the junction of the NWDs for $D_{\text{NW}} > 60$ nm like channel SP mode. Thus, the spectra of SP modes are mainly determined by the structure within the junction and not by the entire structure of the NWDs. Thus, $\lambda_{MAX}$ is insensitive to $D_{NW}$ and maintains at 550 nm for $D_{NW} > 60$ nm, even $\lambda_{LP}$ exhibits redshifts. Indeed, with an increase in $D_{NW}$, the junction structure becomes similar to a V-groove on a plate, where the spectra of SP modes are determined by the taper angle and depth of the V-groove [35]. These properties of $\lambda_{MAX}$ can explain the reason of inconsistency in Fig. 3 as that the spectral shape of $\left| E^{loc}(\lambda_{em}) / E^1(\lambda_{em}) \right|^2$ in Eq. (1) does not change for $D_{NW} > 60$ nm due to the localization of the electric field within the junction of the 1D HSs.

The results shown in Fig. 9 were compared with the experimental results. The relationship between $\lambda_{LPS}$ and $D_{NWS}$ is presented in Fig. 9(d). The dotted lines in Fig. 9(d) indicate the experimentally obtained $\lambda_{LPS}$. The results indicate that the $D_{NWS}$ in the experiment are distributed from ~50 to 100 nm. The relationship between $\lambda_{MAXS}$ and $D_{NWS}$ is shown in Fig. 9(e). The dotted lines in Fig. 9(e) indicate the experimentally obtained $\lambda_{SERRS}$. The dotted lines indicate that the SERRS spectra are distributed within the flat region of $\lambda_{MAXS}$ with respect to $D_{NWS}$. These relationships clearly reveal the spectral region where $\lambda_{MAX}$ maintains 550 nm even $\lambda_{LP}$ increases from 550 to 675 nm by changing $D_{NW}$ from 60 to 120 nm. The spectral region is consistent with the experimental results Fig. 3, which shows the large variations in $\lambda_{LP}$ (525 to 620 nm) and the minor variations in $\lambda_{SERRS}$ (540 to 570 nm). Figure 9(f) presents a plot of the experimentally obtained $\lambda_{LPS}$ with respect to $\lambda_{SERRSS}$ (open red circles) and the plot of the calculated $\lambda_{LPS}$ with respect to $\lambda_{MAXS}$ (open black circles). The calculations well reproduce the experimental property, thus indicating that the minor variations in $\lambda_{SERRSS}$ are due to the tight confinement of the electric fields within junctions and are therefore insensitive to the entire structures of the NWDs. The $\lambda_{MAXS}$ increase again for $D_{NW} > 140$ nm as in Fig. 9(e). This increase may indicate the appearance of a higher-order SP mode due to an increase in the junction height [35].

We found that the SERRS spectra at the edges of 1D HSs exhibited redshifts from the SERRS spectra around the centers of the 1D HSs as shown in Figs. 1 and 3. This property was commonly observed in the SERRS spectra for the narrow- and wide-field excitation. To clarify the redshifts, the position dependence of the spectra of $E_{\text{all}}^{\text{loc}}$ along the 1D HSs was calculated with $D_{\text{NW}} = 60$ nm and $d_{\text{g}} = 0$ nm. Figures 10(a1) and 10(b1) present the $\lambda_{\text{ex}}$ dependence of propagation profiles of $E_{\text{all}}^{\text{loc}}(\lambda_{\text{ex}})$ expressed as contour maps by narrow- and wide-field excitations, respectively. Figures 10(a2) and 10(b2) present the enlarged maps of Figs. 10(a1) and 10(b1), respectively, from $z = -0.1$ to 1 μm. The spectra of $E_{\text{all}}^{\text{loc}}(\lambda_{\text{ex}})$ around the edges and those around the centers look significantly different. Figures 10(a3) and 10(b3) present the spectra of $E_{\text{all}}^{\text{loc}}(\lambda_{\text{ex}})$ at $z = 0.1$ (green curve) and 2.5 μm (red curve) obtained by narrow- and wide-field excitations, respectively. The spectral intensities are normalized by $E_{\text{all}}^{\text{loc}}(\lambda_{\text{ex}})$ at $\lambda_{\text{ex}} = 550$ nm. The spectra of $E_{\text{all}}^{\text{loc}}(\lambda_{\text{ex}})$ at $z = 100$ nm reveal a relatively higher intensity in the spectral regions ~600 nm than those at $z = 2.5$ μm. The difference between the relative intensities is discussed using the spectra of the SP modes supporting the short and long propagations. As the discussion in Fig. 8, the SP modes supporting the short and long propagation are the upper and lower branch of the coupled SP mode. The difference between the two SP modes is in their dephasing rates, which are mainly determined by the ohmic losses of the oscillating electric fields inside the NWs. As observed in Figs. 8(c3a) and 3(c3b), the distribution of $\left| E_{\text{all}}^{\text{loc}} / E_{\text{max}}^{\text{loc}} \right|$ inside the NWD at $z = 100$ nm is larger than that at $z = 2.5$ μm. Thus, the spectrum of SP mode supporting the short propagation is broader than that supporting the long propagation. This broader spectral width is observed as spectral redshifts at the edges of the 1D HSs as shown in Figs. 10(a3) and 10(b3). Indeed, the dispersion relations of coupled SP modes shows that the LDOS spectrum of the upper branch is considerably broader than that of the lower branch in Fig. S3(a3) [24]. In addition, the contribution of the LP resonance, which is not related to the 1D propagation, at the edges needs to be considered. Enhanced electric fields by the LP resonance are observed as spike-type structures at $z = 0$ nm in Fig. 7(a). The excitation and SERRS light are efficiently coupled with and scattered by the LP resonance, respectively, as Eq. (1). Thus, the spectrum of LP resonance influences the redshifts of SERRS spectra at the edges of the 1D HSs. Figures 10(a2) and 10(b2) present such LP resonance peaks at approximately 650 nm at $z = 0$ nm. We consider that these peaks also contribute to the observed redshifts in SERRS spectra.

We compared the calculation results with the experimental results to confirm that the redshifts are induced by differences in the spectral broadness of the SP modes supporting the short and long propagations. The ratios between the SERRS intensities at $\lambda_{ex} = 550$ nm and those at 600 nm at the edges are plotted with respect to the ratios at the centers in Fig. 4(c). The calculated ratios are added using the spectra of $\left| E_{all}^{loc} \right|^2$ as $\left| E_{all}^{loc} \left( 550 \text{nm} \right) / E_{all}^{loc} \left( 600 \text{nm} \right) \right|^2$ for the narrow- (blue-open circles) and wide-field excitation (red-open circles), as shown in Fig. 10(c). The values of $D_{NW}$ are set as 40, 60, and 80 nm regarding the estimated $D_{NW}$ in Fig. 9(d). The calculated distribution of ratios is consistent with the experimental results, thus indicating that the observed redshifts in the SERRS spectra are induced by the spectral difference between the SP modes supporting the short and long propagation of SERRS light.

We observed the large NWD-by-NWD variation in $L_p$ of SERRS light along the 1D HS in Fig. 5. The value of $L_p$ decreases by the losses arising from scattering or absorption by e.g. defects [26]. Thus, the variation in $L_p$ may be mainly attributed to the degree of these factors. We here discuss another factor arising the variation in $L_p$ using the propagation profiles of $E_{all}^{loc}$. We focus on the long propagation profiles. Figures 8(a3)–8(a5) and 9(a1)–9(a3) indicate that the $\lambda_{ex}$ showing a channel-type propagation profile of $E_{all}^{loc}$ is dependent on both $d_g$ and $D_{NW}$. These dependences indicate that $E_{all}^{loc}$ immediately decays during propagation if $\lambda_{ex}$ deviates from the $\lambda_{MAX}$ of SP modes supporting the long propagation. Thus, the degree of deviation may be another factor arising the variations in $L_p$. In other words, the mismatch between $\lambda_{ex}$ and $\lambda_{MAX}$ induces the variation in $L_p$s. Thus, we examined the $d_g$ and $D_{NW}$ dependences of propagation profiles of $E_{all}^{loc}$. Figures 11(a1)–11(c1) present the $\lambda_{ex}$ dependences of propagation profiles of $E_{all}^{loc}\left(\lambda_{ex}\right)$ for $d_g$ = -2, -1, and 0 nm, respectively, for $D_{NW}$ = 60 nm. Figures 11(c1)–11(e1) present the propagation profiles of $E_{all}\left(\lambda_{ex}\right)$ for $D_{NW}$ = 60, 40, and 30 nm, respectively, for $d_g$ = 0 nm. The $\lambda_{MAX}$s redshift from 500 to 550 nm with an increase in $d_g$ as in Figs. 11(a1)–11(c1). The $\lambda_{MAX}$s blueshift from 550 to 475 nm with a decrease in $D_{NW}$ as in Figs. 11(c1)–11(e1). The propagation profiles of $\left|E_{all}^{loc}/E_{max}^{loc}\right|$ at $\lambda_{ex}$ = 550 nm are extracted from Figs 11(a1)–11(e1) as in Figs 11(a2)–11(e2). Notably, Figs. 11(a2)–11(e2) reveal that $E_{all}^{loc}$ immediately decays within 0.5 to 2 μm if $\lambda_{ex}$ deviates from $\lambda_{MAX}$. This fact indicates that the experimentally observed NWD-by-NWD variations in $L_p$ in Fig. 5 are partially induced by the mismatch between $\lambda_{ex}$ and $\lambda_{MAX}$. We check the effect of the length of the 1D HSs on $L_p$, because the reflection from another edge may modulate the propagation profiles of $E_{all}^{loc}$ and confuse the discussion. Figures 11(f1)–11(f3) present the $\lambda_{ex}$ dependences of propagation profiles with $D_{NW}$ = 60 nm and $d_g$ = 0 nm for three 1D HSs with lengths of 4.5, 3.5, and 2.5 μm, respectively. Their propagation profiles are almost identical until the edges even changing the lengths, thus indicating that the effect of reflection is negligible. Thus, the propagation profiles for the short 1D HSs are simply estimated by cutting the propagation profiles of the long 1D HS at the $z$ position of the short 1D HS. Thus, we can compare the experimentally obtained variations in $L_p$ having various HS lengths with the calculated $L_p$s having a certain HS length of 4.5 μm. The propagation profile of Fig. 11(c2) is employed as the standard for determining the value of $L_p$. The $L_p$ is determined as the value of $z$, where the normalized amplitude of $\left| E_{all}^{loc} / E_{max}^{loc} \right|$ (or the normalized intensity of $\left| E_{all}^{loc} / E_{max}^{loc} \right|^2$) is ~17% (or ~3%) of the maximum. Note that the definition of $L_p$ was already employed in the experimental results in Fig. 5(a). Figure 11(g) presents the plots of the expected $L_p$ vs. HS lengths using Fig. 11(a2)–11(e2) with the experimentally obtained $L_p$s in Fig. 5(b). The NWD-by-NWD variations in the experimentally obtained $L_p$s are consistent with the calculated variations, thus indicating that the match between $\lambda_{ex}$ and $\lambda_{MAX}$ is critical for maximizing $L_p$.

## IV. CONCLUSIONS

In this study, we evaluated the experimentally observed propagation of SERRS light through 1D HSs located between plasmonic NWDs by FDTD calculations. The observed properties are as follows. The SERRS light polarized orthogonal to the NWD long axis effectively propagates along 1D HSs even excitation polarization parallel to the NWD long axis. There are two types of propagation along 1D HSs. One is the short propagation with high intensity, and another is the long propagation with low intensity. The SERRS spectra are independent of the LP resonance spectra of NWDs. Furthermore, the SERRS spectra of 1D HS edges are redshifted from the SERRS spectra of center positions. There are significant variations in the propagation lengths between 1D HSs. These properties were well reproduced by the FDTD calculations of NWDs with $D_{NW} = 60$ nm and $d_g = 0$ nm, which proposed the following propagation mechanism: Excitation light is resonantly coupled with LP at the edge of HS, and the light energy is provided to two types of SP modes supporting the short and long propagation. The SP mode supporting short and long propagation are likely the upper and lower branch of the coupled SP modes between the DD- and QQ-coupled mode. These coupled SP modes are confirmed by the dispersion relations [24]. The spectral matching between $\lambda_{ex}$ and $\lambda_{MAX}$ is critical for maximizing the value of $L_p$.

Finally, we discuss the potential applications of the propagation through 1D HSs. LPs and molecular excitons strongly interact at the edges of 1D HSs [10]. Thus, 1D HSs will be useful for the remote excitation and detection of various phenomena related to strong light–matter interactions. Such phenomena include strong coupling, in which the EM coupling rates between LP and molecular excitons are larger than the dephasing rates of both LP and molecular excitons [12,16,17,37], and ultra-fast surface-enhanced fluorescence (ultra-fast SEF), in which the SEF rates exceed the molecular vibrational decay rates, thus resulting in emissions from vibrationally excited states in an electronically excited state [36,38], and others [39]. In particular, strong coupling under near-single-molecule conditions will be a major topic in polariton chemistry, which covers the research fields of modified photoemission, changes in excited-state quantum yields, effects on photochemical reactions, vibrational strong coupling, effects on the ground state, intermolecular communications, and energy transfer, among others [12,40]. The 1D HSs can serve as a platform for the remote excitation and detection of various phenomena in polariton chemistry. Such platforms serve as a basis for new research fields by connecting cavity quantum electrodynamics and plasmon-enhanced spectroscopy [12,40,41].

**ACKNOWLEDGMENTS**

This study was supported in part by KIBAN C [grant number 18K04988] from the Ministry of Education, Culture, Sports, Science, and Technology of Japan.

## Figure captions

FIG. 1 (a1) Experimental setup for SERRS imaging by wide-field excitation and narrow-field excitation for a single NWD. Inset presents the focusing laser beam spot. The NWDs were placed on the cover glass. (b1) and (b2) 1D propagation measured by narrow-field excitation for single NW with laser light polarized orthogonally and parallelly, respectively, to the long axis of the NW. (c1)–(c4) Dark-field images of single NWDs measured by white-light excitation; (d1)– (d4) SERRS images of single NWDs measured by wide-field excitation using circularly polarized laser light; (e1)–(e4) SERRS images of single NWDs measured by narrow-field excitation using laser light polarized orthogonally to NWD long axis; and (f1)–(f4) SERRS images of single NWDs measured by narrow-field excitation using laser light polarized parallel to NWD long axis. Scale bars are commonly 5 μm.

FIG. 2 (a1) Dark-field images from white-light excitation of single NWDs. The SERRS images of the single NWD measured by wide-field excitation using circularly polarized laser light and detection of (a2) orthogonally and (a3) parallelly polarized SERRS light with respect to the NWD long axis; SERRS images of single NWD measured by narrow-field excitation using (b1) orthogonally and (b2) parallelly polarized light with respect to the NWD long axis; (b3) SERRS intensities with respect to polarization angle; SERRS images of single NWDs measured by narrow-field excitation using (c1) parallelly and (c2) orthogonally polarized light with respect to the NWD long axis; and (c3) SERRS intensities with respect to the polarization angle. Scale bars are commonly 5 μm.

FIG. 3 (a1)–(a3) Spectra of the elastic light scattering for three NWDs. LP resonance peaks are indicated by blue arrows. (b1)–(b3) Spectra of the SERRS intensity of the same NWDs measured by wide-field excitation using circularly polarized laser light. Peaks of envelopes of SERRS spectra are indicated by red arrows. (c) Relationship between $\lambda_{\text{LPS}}$ and $\lambda_{\text{SERRSS}}$.

FIG. 4 The SERRS images of the single NWDs measured by (a1) wide-field excitation using circularly polarized laser light and (a2) narrow-field excitation using orthogonally polarized light

with respect to the NWD long axis. The redshifted colors of the edge of the 1D HS are indicated by a white dotted open circle. Scale bars in (a1) and (a2) are 5.0 μm. (b) Spectra of the SERRS intensity from the center (green curve) and edge (red curve) of the 1D HS. The intensities at 550 nm and 600 nm are indicated by the green and red arrows, respectively. (c) Ratios between SERRS intensities at 550 nm and 600 nm at the edges of the 1D HSs with respect to the ratios at the centers of the 1D HSs for 20 NWDs.

FIG. 5 (a1)–(a3) Propagation profiles of SERRS light and entire profiles of 1D HSs for three NWDs measured by wide-field excitation (dotted lines) and narrow-field excitation (solid lines). Side panels indicate SERRS images measured by wide-field excitation (blue frame) and narrow-field excitation (red frame). $L_p$ is defined as the distance at which the SERRS intensity is 3% of the maximum SERRS intensity from the excitation spot, as indicated by red arrows. The entire length of the 1D HS is defined as the distance from the edge to the edge of the SERRS images measured by wide-field excitation, as indicated by blue arrows. (b) Relationship between entire lengths of 1D HSs and $L_p$s for 30 NWDs.

FIG. 6 (a) FDTD calculation setup for electric fields around single NWD calculated by wide-field excitation (from upper side) and narrow-field excitation (from bottom side) with the coordinate system. (b) Spectra of the elastic light scattering for the NWD calculated by wide-field excitation (red curve) for the incident light polarized with respect to the x-axis and for the incident light polarized with respect to the z-axis (black curve). (c1)–(e1) Images of the NWD cross-sections in the x-y, x-z, and y-z planes, respectively. Red and black arrows indicate the directions of polarization. (c2) and (c3) $E_{all}$ distributions of the x-y plane at $z = 2.5$ μm and $\lambda_{ex} =$ 550 nm for the incident light polarized with respect to the x- and z-axis, respectively. (d2) and (d3) $E_{\text{all}}$ distributions of x-z plane at $y = $ -9 nm and $\lambda_{\text{ex}} = 550$ nm for the incident light polarized with respect to the x- and z-axis, respectively. (e2) and (e3) $E_{\text{all}}$ distributions of y-z plane at $x = 0$ nm and $\lambda_{\text{ex}} = 550$ nm for the incident light polarized with respect to the x- and z-axis.

FIG. 7 FDTD calculations of propagation profiles along a 1D HS of the NWD with $D_{\text{NW}} = 60$ nm and $d_{\text{g}} = 0$ nm under Cross–Nicole conditions. (a1)–(a3) Propagation profiles of $E_{\text{all}}^{\text{loc}}$, $E_{\text{x}}^{\text{loc}}$, and $E_{\text{z}}^{\text{loc}}$ at y = -9 nm, respectively, for the excitation polarization orthogonal to the NWD long axis at $\lambda_{\text{ex}} = 550$ nm. (b1)–(b3) $\lambda_{\text{ex}}$ dependence of the propagation profiles of $E_{\text{all}}^{\text{loc}}$, $E_{\text{x}}^{\text{loc}}$, and $E_{\text{z}}^{\text{loc}}$ at y = -9 nm, respectively, expressed as contour maps, for the excitation polarization orthogonal to the NWD long axis. The red and blue frames in (a) correspond to the two contour maps with red and blue frames in (b), respectively. (c1)–(c3) Propagation profiles of $E_{\text{all}}^{\text{loc}}$, $E_{\text{x}}^{\text{loc}}$, and $E_{\text{z}}^{\text{loc}}$ at y = -9 nm, respectively, for the excitation polarization parallel to the 1D HS at $\lambda_{\text{ex}} = 550$ nm. (d1)–(d3) $\lambda_{\text{ex}}$ dependence of the propagation profiles of $E_{\text{all}}^{\text{loc}}$, $E_{\text{x}}^{\text{loc}}$, and $E_{\text{z}}^{\text{loc}}$ at y = -9 nm, respectively, expressed as contour maps, for the excitation polarization parallel to the NWD long axis. The red and blue frames in (c) correspond to the two contour maps with red and blue frames in (d), respectively. The white color in the contour maps in red (or blue) frames indicates that the EF is smaller (or larger) than the minimum (or maximum) of the scale bar.

FIG. 8 FDTD calculations of $d_{\text{g}}$ dependence of propagation profiles along a 1D HS of the NWD with excitation polarization orthogonal to the NWD long axis for $D_{\text{NW}} = 60$ nm. (a1)–(a5) $\lambda_{\text{ex}}$ dependence of the propagation profiles of $E_{\text{all}}^{\text{loc}}$ expressed as contour maps for $d_{\text{g}} = 20, 5, 0, -1,$

and -5 nm, respectively. The white color in the contour maps around $z = 0$ indicates that the EF is larger than the maximum of the scale bar. (b1)–(b5) $\sigma_{sca}$ spectra of wide-field excitation (black curves) and $\lambda_{ex}$ dependence of $E_{all}^{loc}$ at $z = 2.5$ μm (red curves) for $d_g = 20, 5, 0, -1,$ and -5 nm, respectively. The white color at approximately z = 0 nm in the contour maps indicate that EFs are larger than the maximum of the scale bar. (c1a)–(c5a) Cross-sections of $\left| E_{all}^{loc} / E_{max}^{loc} \right|$ of x-y plane at $z = 100$ nm (red frames). (c1b)–(c5b) Cross-sections of $\left| E_{all}^{loc} / E_{max}^{loc} \right|$ of x-y plane at $z = 2.5$ μm (black frames). The insets of (c3a) and (c3b) present the enlarged cross-sections around junctions. The scale bars of the insets in (c3a) and (c3b) are 5 nm. (d) Relationship between $d_g$s and oscillation periods (open circle), and relationship between $d_g$s and decay lengths up to 15% of the maximum amplitude (open triangle). (e) Relationship between $d_g$s and $\lambda_{LPS}$. The dotted lines indicate the experimentally obtained $\lambda_{LPS}$ shown in Fig. 3(c). (e) Relationship between $d_g$s and $\lambda_{MAXS}$. The dotted lines indicate the experimentally obtained $\lambda_{SERRS}$s shown in Fig. 3(c).

FIG. 9 FDTD calculations of $D_{NW}$ dependence of propagation profiles along a 1D HS with excitation polarization orthogonal to the NWD long axis for $d_g = 0$ nm. (a1)–(a5) $\lambda_{ex}$ dependence of the propagation profiles of $E_{all}^{loc}$ expressed as contour maps for $D_{NW} = 20, 40, 60, 80,$ and 100 nm, respectively. The white color around $z = 0$ nm in the contour maps indicate that EFs are larger than the maximum of the scale bar. (b1)–(b5) $\sigma_{sca}$ spectra of wide-field excitation (black curves) and $\lambda_{ex}$ dependence of $E_{all}^{loc}$ at $z = 2.5$ μm (red curves) for $D_{NW} = 20, 40, 60, 80,$ and 100 nm, respectively. (c1a)–(c5a) Cross-sections of $\left| E_{all}^{loc} / E_{max}^{loc} \right|$ of x-y plane at $z = 100$ nm (red frames). (c1b)–(c5b) Cross-sections of $\left| E_{all}^{loc} / E_{max}^{loc} \right|$ of x-y plane at $z = 2.5$ μm (black frames). (d)

Relationship between $D_{\text{NWS}}$ and $\lambda_{\text{LPS}}$. Dipole and quadrupole plasmon mode indicated by open circles and triangles, respectively. (e) Relationship between $D_{\text{NWS}}$ and $\lambda_{\text{MAXS}}$. The dotted lines indicate the experimentally obtained $\lambda_p$ values shown in Fig. 3(c). (f) Relationship between $\lambda_{\text{LPS}}$ and $\lambda_{\text{MAXS}}$. The dotted lines indicate the experimentally obtained $\lambda_{\text{SERRSS}}$ shown in Fig. 3(c).

FIG. 10 FDTD calculations for the position dependence of the $E_{\text{all}}^{\text{loc}}\left(\lambda_{\text{ex}}\right)$ spectra along a 1D HS with excitation polarization orthogonal to the NWD long axis and $D_{\text{NW}} = 60$ nm. (a1) and (a2) $\lambda_{\text{ex}}$ dependence of the propagation profiles of $E_{\text{all}}^{\text{loc}}$ expressed as contour maps calculated by narrow-field excitation (red frame) and its enlarged contour maps (green frame) from -0.1 to 1 μm, respectively. The white color in (a1) around $z = 0$ nm in the contour maps indicates EFs larger than the maximum of the scale bar. The white color in (a2) indicates that the EF is smaller than the minimum of the scale bar. (a3) $\lambda_{\text{ex}}$ dependence of $E_{\text{all}}^{\text{loc}}$ at $z = 100$ nm (green curves) and 2.5 μm (red curves) of (a1) and (a2), respectively. (b1) and (b2) $\lambda_{\text{ex}}$ dependence of the propagation profiles of $E_{\text{all}}^{\text{loc}}$ expressed as contour maps calculated by wide-field excitation (red frame) and the enlarged contour maps (green frame) from $z = -0.1$ to 1 μm, respectively. (b3) $\lambda_{\text{ex}}$ dependence of $E_{\text{all}}^{\text{loc}}$ at $z = 100$ nm (green curves) and 2.5 μm (red curves) of (a1) and (a2), respectively. (f) Relationship between $\left|E_{\text{all}}^{\text{loc}}\left(550\text{nm}\right)/E_{\text{all}}^{\text{loc}}\left(600\text{nm}\right)\right|^2$ at $z = 100$ nm and those of $z = 2.5$ μm by narrow-field excitation (blue open circles) and wide-field excitation (red open circles). The relationship is plotted in Fig. 4(c).

FIG. 11 FDTD calculations for $L_p$ of $E_{all}^{loc}(\lambda_{ex})$ along 1D HSs with excitation polarization orthogonal to the NWD long axis. (a1)–(b1) $\lambda_{ex}$ dependence of the propagation profiles of $E_{all}^{loc}$ expressed as contour maps for $d_g$ = -1 and -2 nm, respectively, with $D_{NW}$ = 60 nm. (c1)–(e1) The $\lambda_{ex}$ dependence of the propagation profiles of $E_{all}^{loc}$ expressed as contour maps for $D_{NW}$ = 60, 40, and 30 nm, respectively, with $d_g$ = 0 nm. (a2)–(b2) Propagation profiles of $E_{all}$ by narrow-field excitation (black curve) and wide-field excitation (blue curve) at $\lambda_{ex}$ = 550 nm for $d_g$ = -1 nm and -2 nm, respectively, with $D_{NW}$ = 60 nm. (c2)–(e2) Propagation profiles of $E_{all}^{loc}$ by narrow-field excitation (black curve) and wide-field excitation (blue curve) at $\lambda_{ex}$ = 550 nm for $D_{NW}$ = 60, 40, 30 nm, respectively, with $d_g$ = 0 nm. (f1)–(f3) $\lambda_{ex}$ dependence of the propagation profiles of $E_{all}^{loc}$ captured by narrow-field excitation expressed as contour maps for1D HSs with lengths of 4.5, 3.5, and 2.5 μm, respectively, and for $D_{NW}$ = 60 nm and $d_g$ = 0 nm. (g) Relationship between $L_p$s and lengths of 1D HS obtained using (a2)–(e2). The open circles indicate the experimentally obtained relationship in Fig. 5(b). The white color around z = 0 in the contour maps indicate EFs larger than the maxima of the scale bars.

FIG. S1 (a1)–(a4) Dark-field images of single NWDs measured by white-light excitation. (b1)–(b4) SERRS images of single NWDs of (a1)–(a4) measured by wide-field excitation using circularly polarized laser light. (c1)–(c4) Scanning electron microscopy (SEM) images of single NWDs obtained from the same area of (a1)–(a4). Scale bars in (b)–(e) are commonly 5 μm.

FIG. S2 (a1) FDTD calculation setup for electric fields around single NW by wide-field excitation (from upper side) and narrow-field excitation (from bottom side). (b) Calculated spectra of the elastic light scattering for the NW by wide-field excitation for the incident light polarized with respect to the x-axis (red curve) and z-axis (black curve), respectively. (c1)–(e1) Images of the NW cross-section in the x-y, x-z, and y-z planes, respectively. Red and black arrows indicate the direction of the polarization. (c2a) and (c2b) $E_z^{loc}$ and $E_x^{loc}$ distributions of x-y plane at $z = 2.5$ μm and $\lambda_{ex} = 550$ nm for the incident light polarized with respect to the z-axis. (c3a) and (c3b) $E_z^{loc}$ and $E_x^{loc}$ distributions of x-y plane at $z = 2.5$ μm and $\lambda_{ex} = 550$ nm for the incident light polarized with respect to the x-axis. (d2)–(d3) $E_{all}^{loc}$ distributions of x-z plane at $y = 0$ nm and $\lambda_{ex} = 550$ nm for the incident light polarized with respect to the z- and x-axis, respectively. (e2)–(e3) $E_{all}^{loc}$ distributions of y-z plane at $x = 0$ nm and $\lambda_{ex} = 550$ nm for the incident light polarized with respect to z- and x-axis, respectively. (f1) and (f2) $\lambda_{ex}$ dependence of the propagation profiles of $E_{all}^{loc}$ by narrow-field excitation expressed as contour maps for the incident light polarized with respect to x- and z-axis, respectively. The observation point of the electric fields is indicated by a black point. (g1) and (g2) Propagation profiles of $E_{all}^{loc}$ (black curve), $E_z^{loc}$ (blue curve), and $E_x^{loc}$ (red curve) by narrow-field excitation at $\lambda_{ex} = 550$ nm for the incident light polarized with respect to the x- and z-axis, respectively. The observation point of the electric fields is indicated by a black point.

FIG. S3 (a1)–(a4) Dispersion relations of LDOS for SP modes of non-touching (a1) and (a2) and overlapping (a3) and (a4) NWDs. The gap distances $d_g$ is shown in the panels. LDOS outside at the gap center in (a1) and (a2) and at a point situated 10 nm outside the NWDs neck in (a3) and (a4). We modify Fig. 3 of [24] for SP mode analysis. The classification of SP modes is based on [23]. (b) Propagation profiles of $E_{\text{all}}^{\text{loc}}$ by narrow-field excitation at $\lambda_{\text{ex}} = 550$ nm for $d_{\text{g}} = 20$ to 0 nm and -2 nm, respectively, with $D_{\text{NW}} = 60$ nm.

Fig. 1

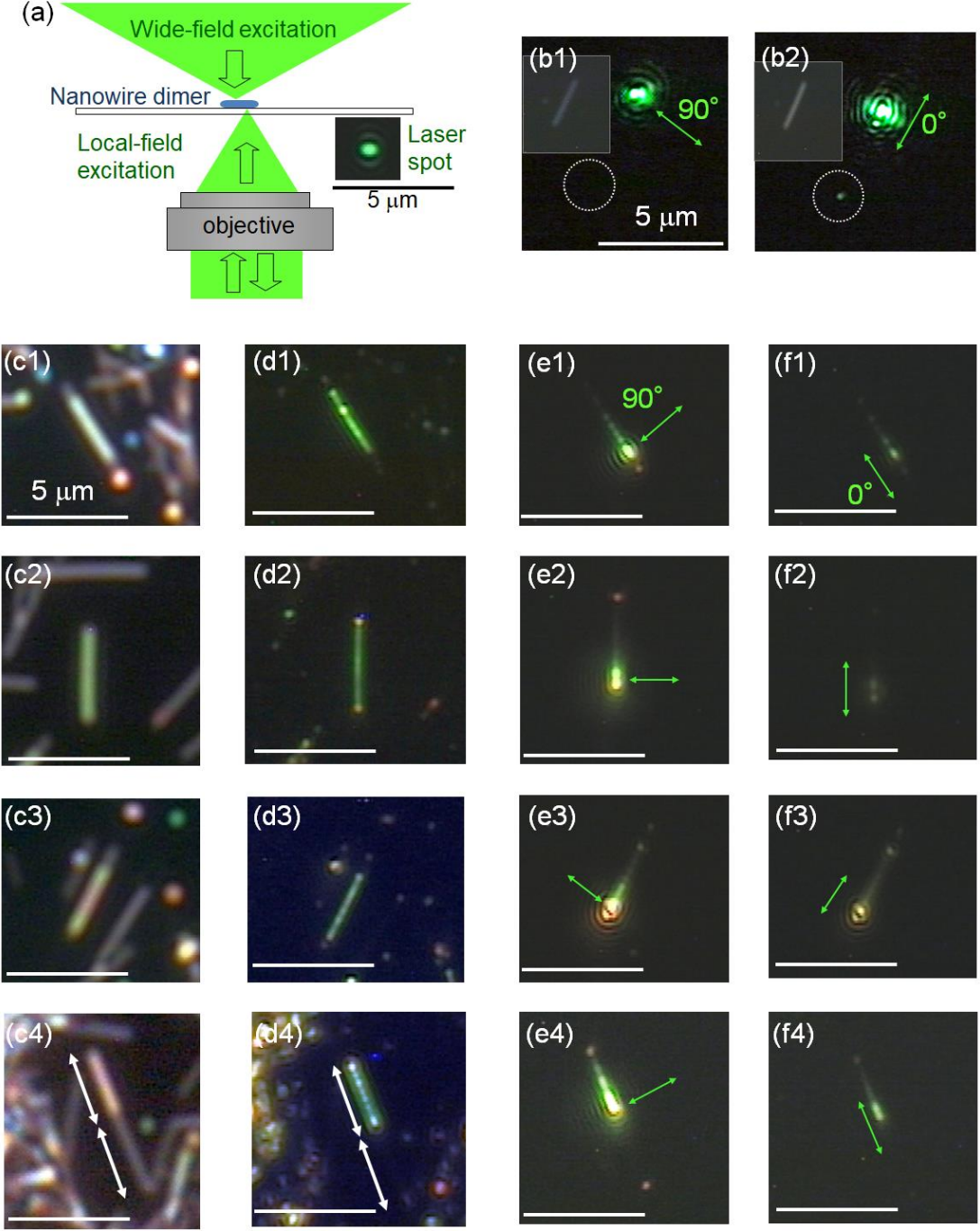

Fig. 2

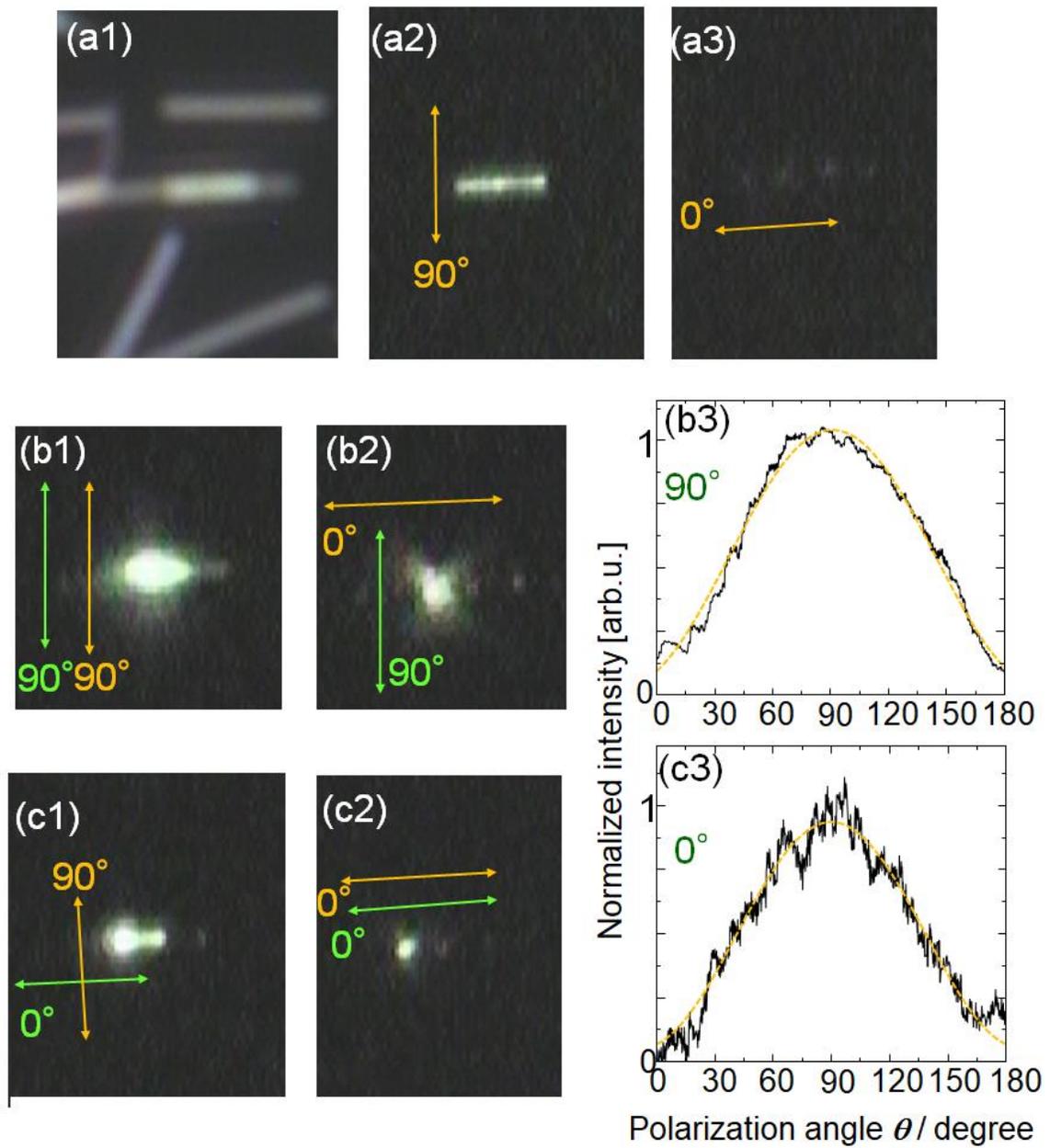

Fig. 3

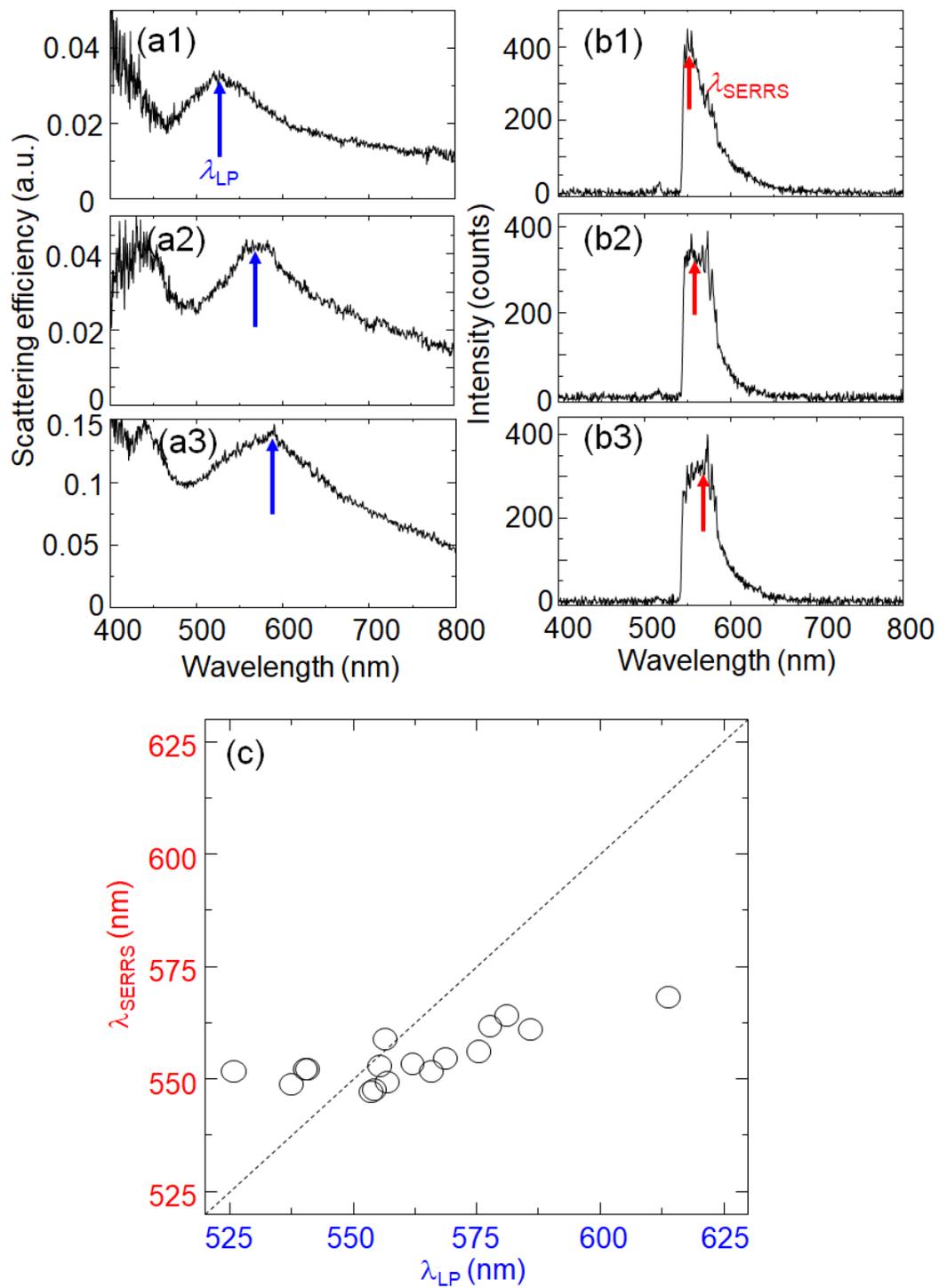

Fig. 4

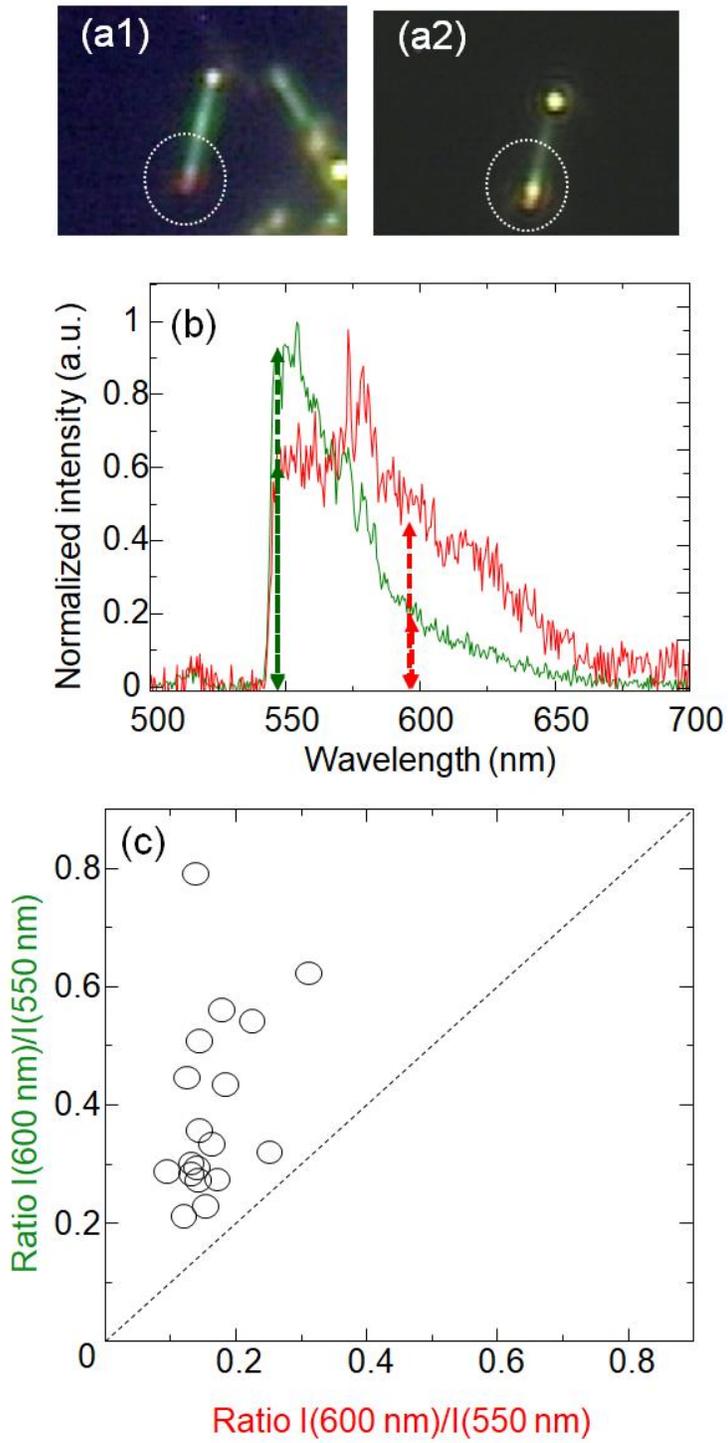

Fig. 5

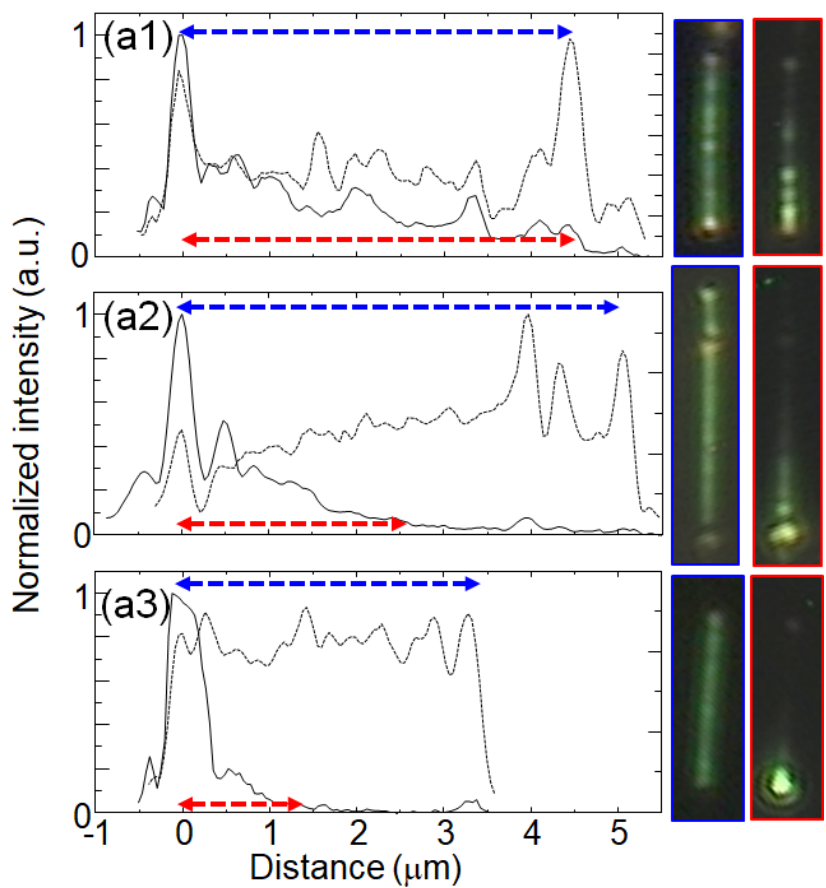

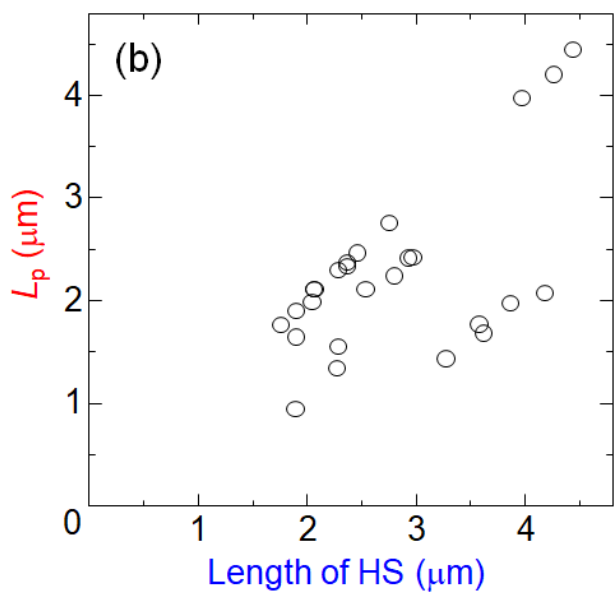

Fig. 6

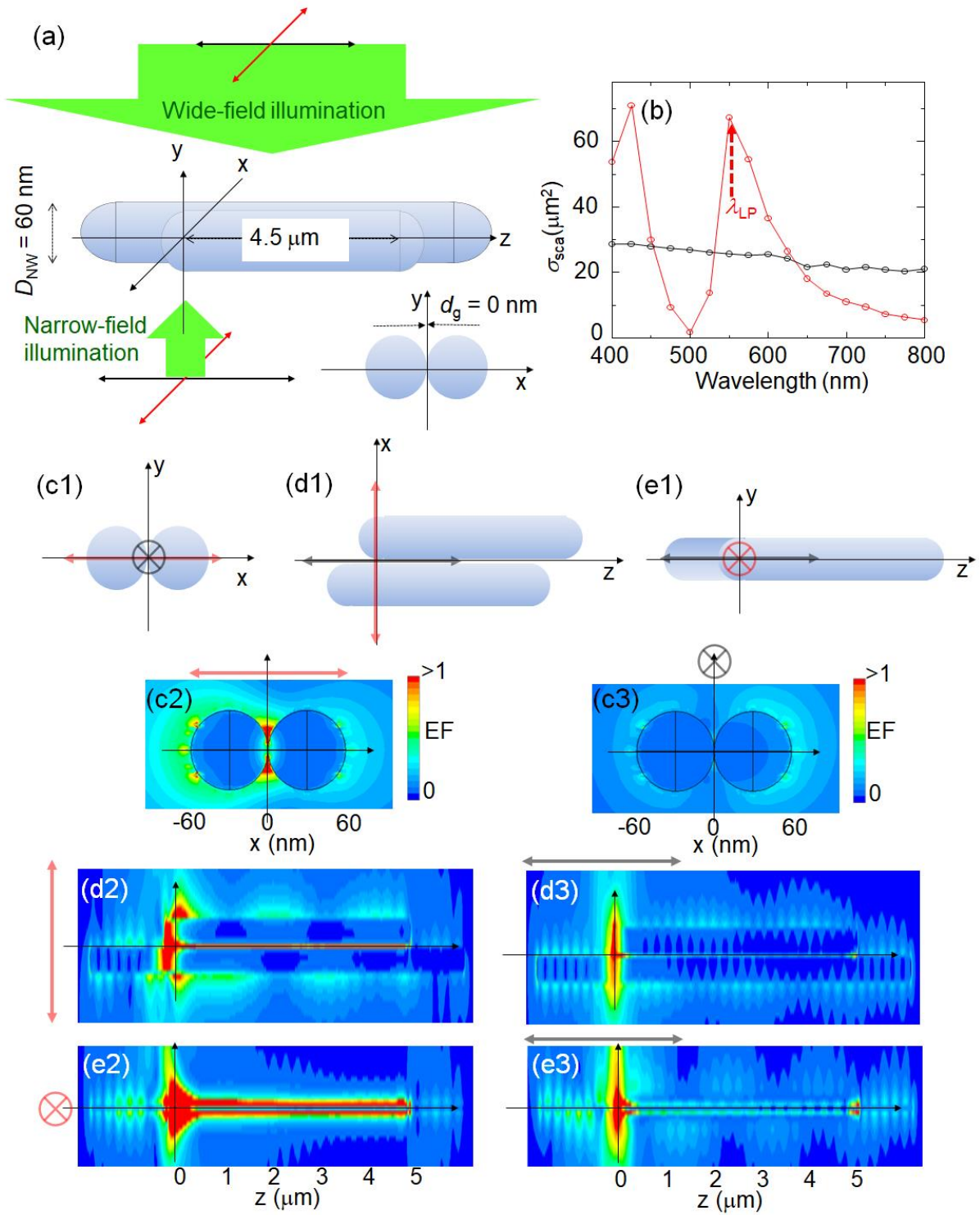

(a) Wide-field illumination

$D_{NW} = 60$ nm    4.5 μm

Narrow-field illumination

$d_g = 0$ nm

(b)

$\lambda_{LP}$

(c1)  (d1)  (e1)

(c2)  EF  (c3)  EF

-60    0    60    -60    0    60
x (nm)    x (nm)

(d2)  (d3)

(e2)  (e3)

0  1  2  3  4  5    0  1  2  3  4  5
z (μm)    z (μm)

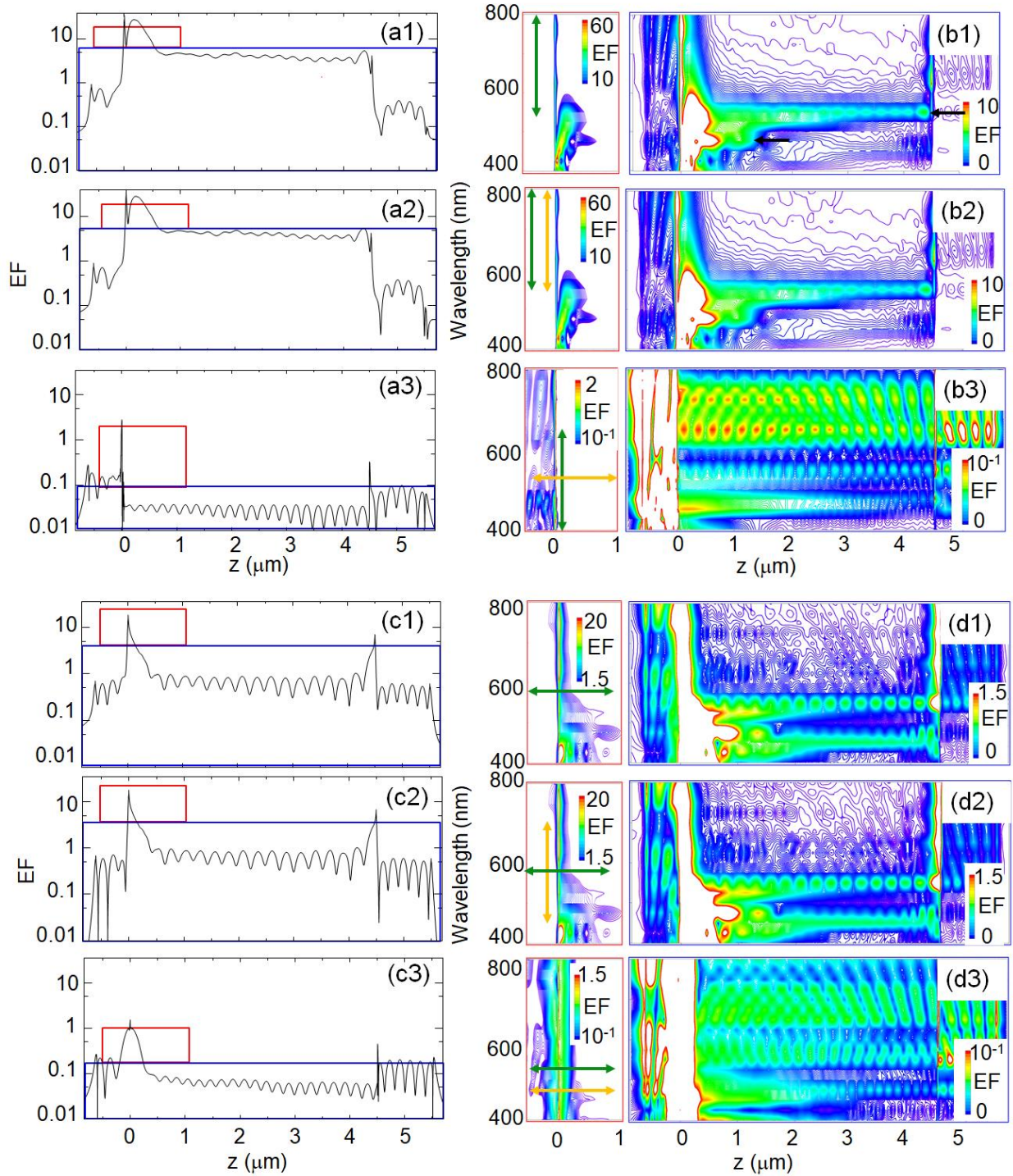

Fig. 8

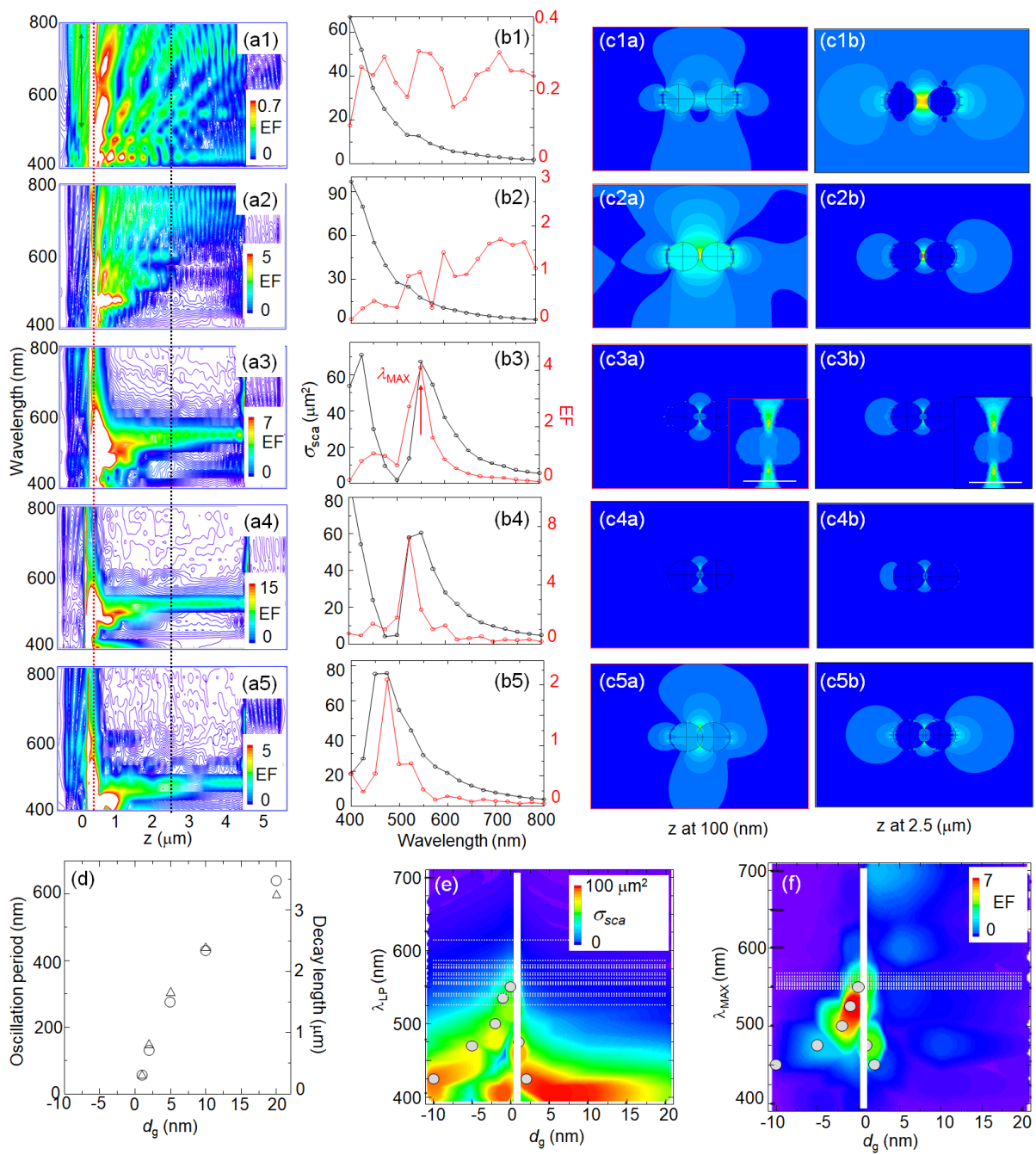

Fig. 9

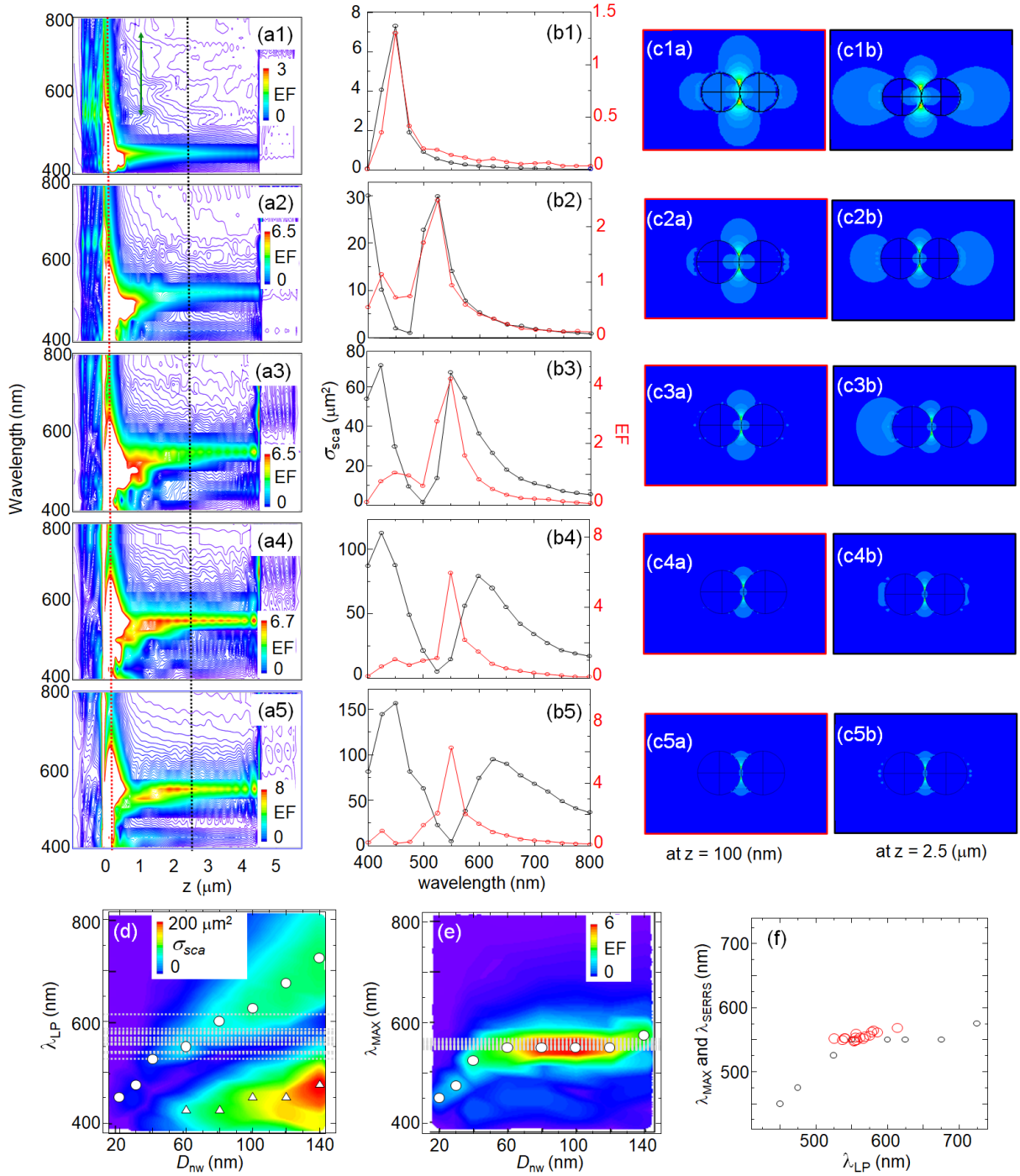

Fig. 10

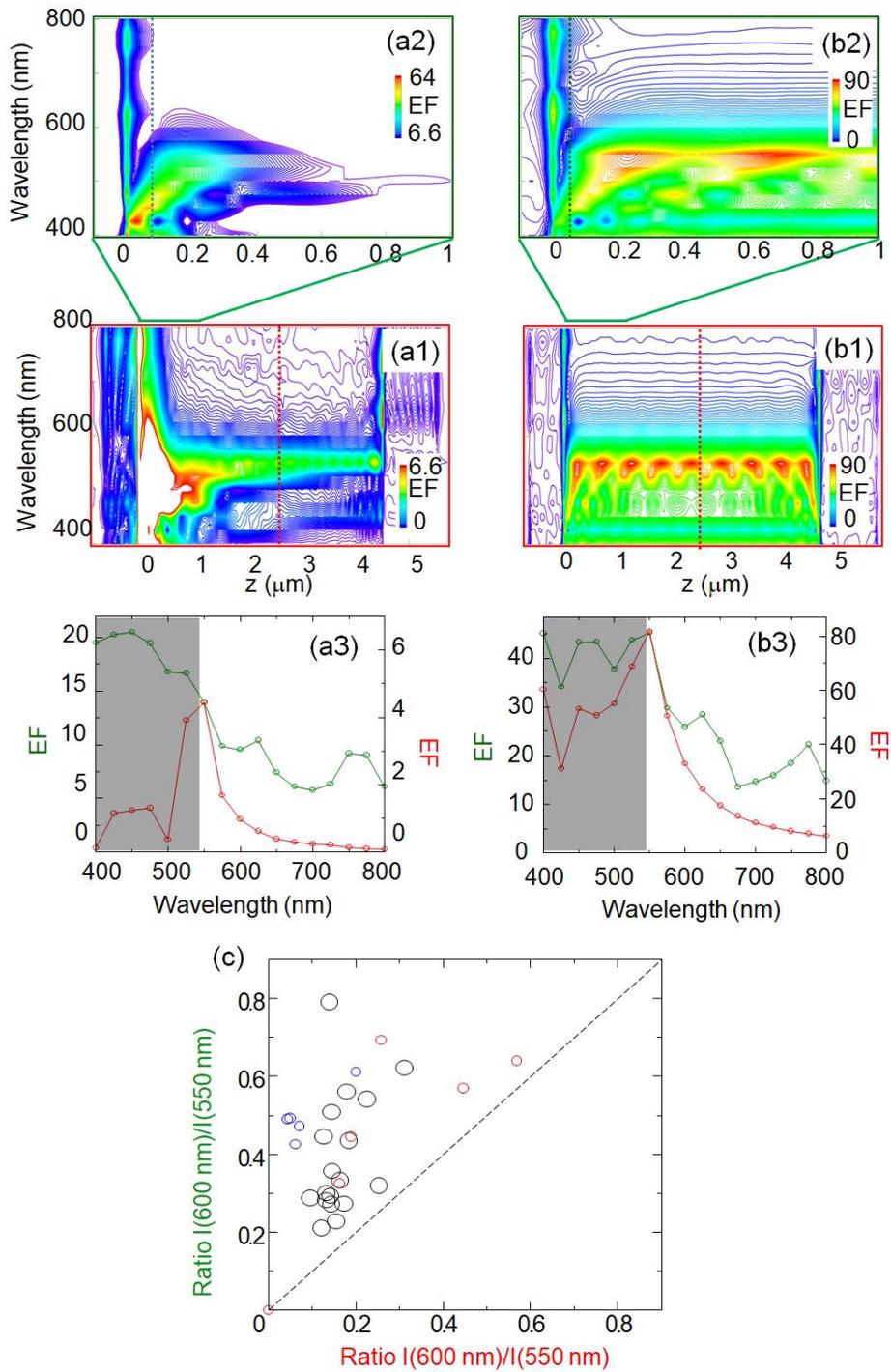

Fig. 11

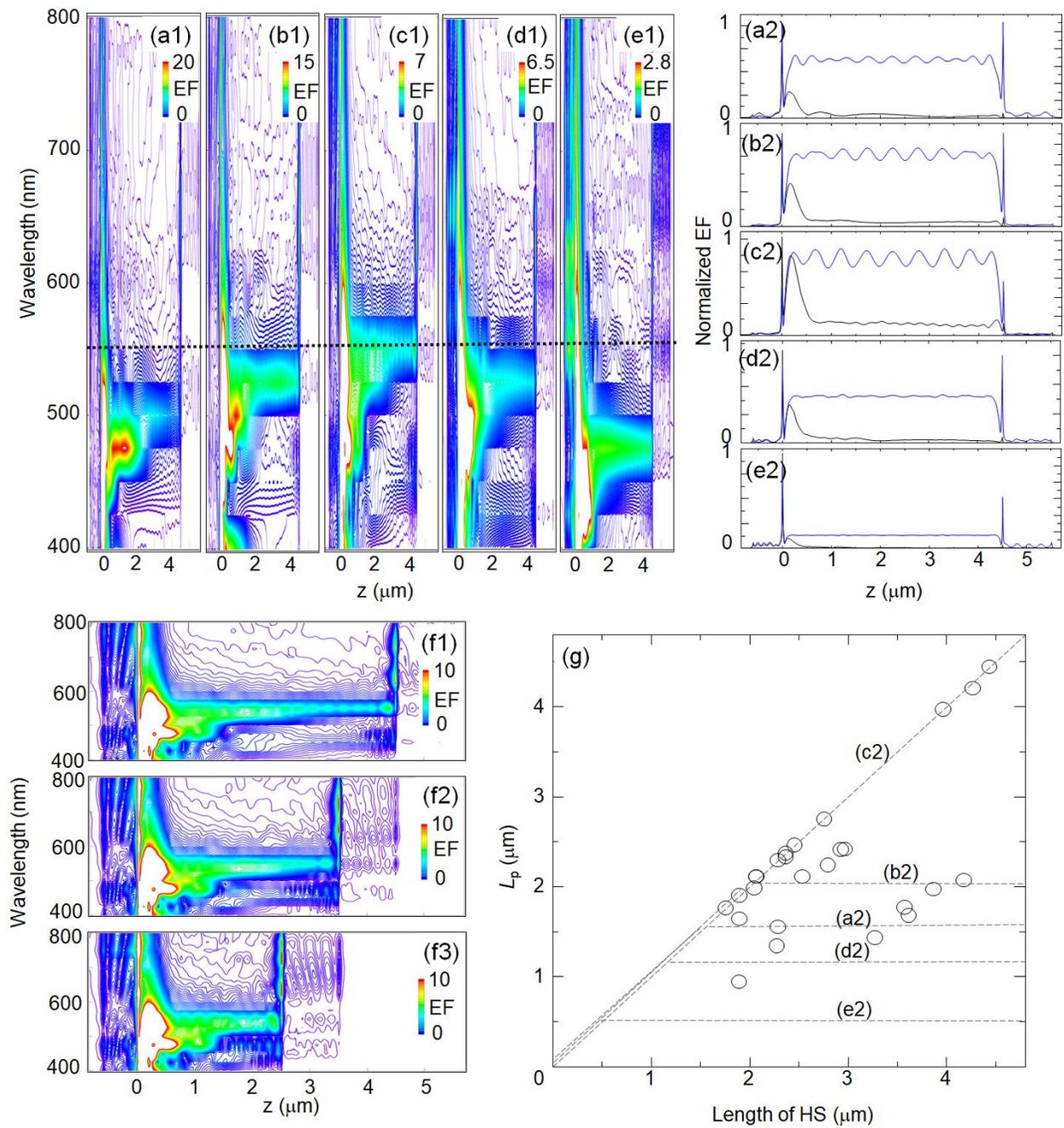

Fig. S1

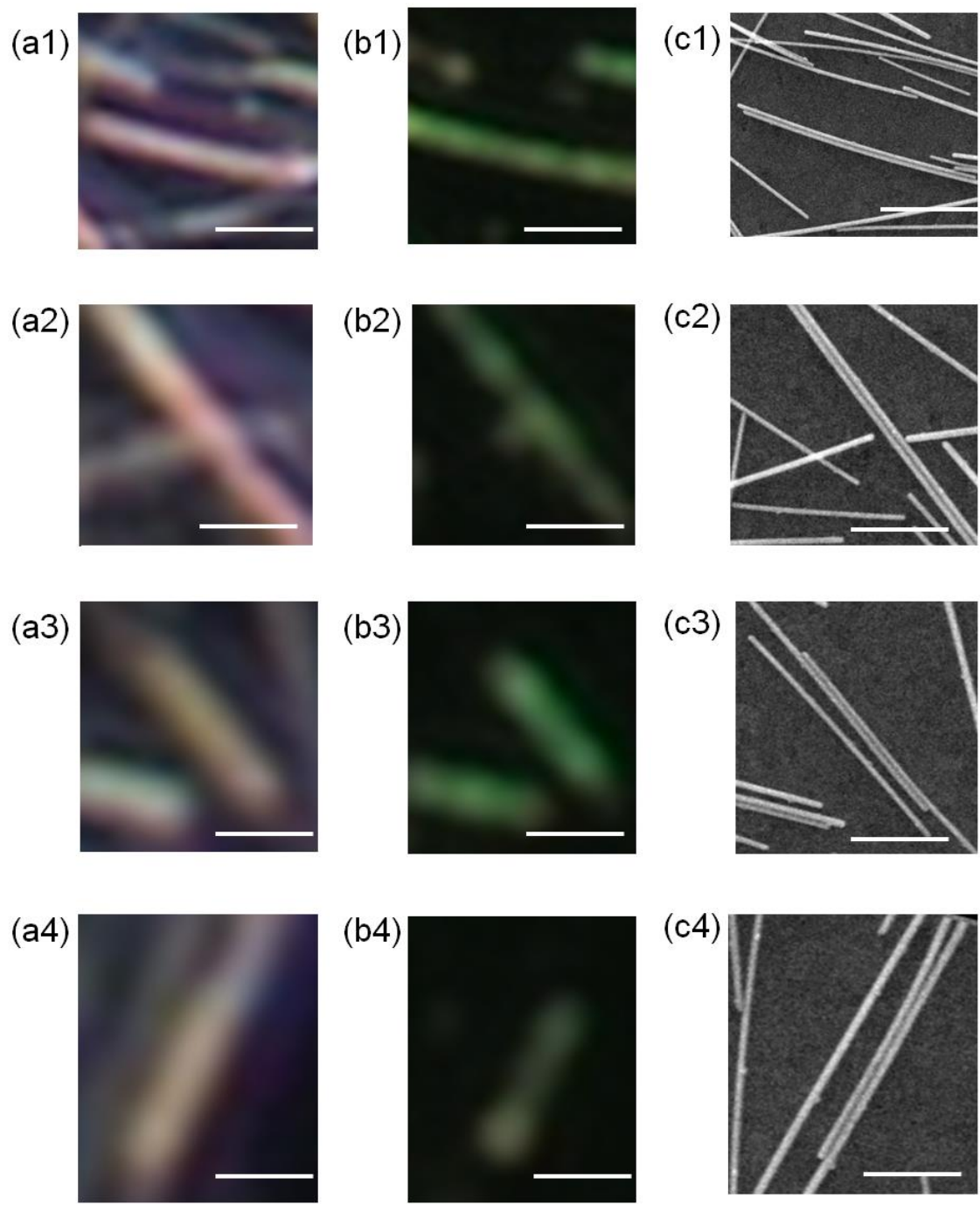

Fig. S2

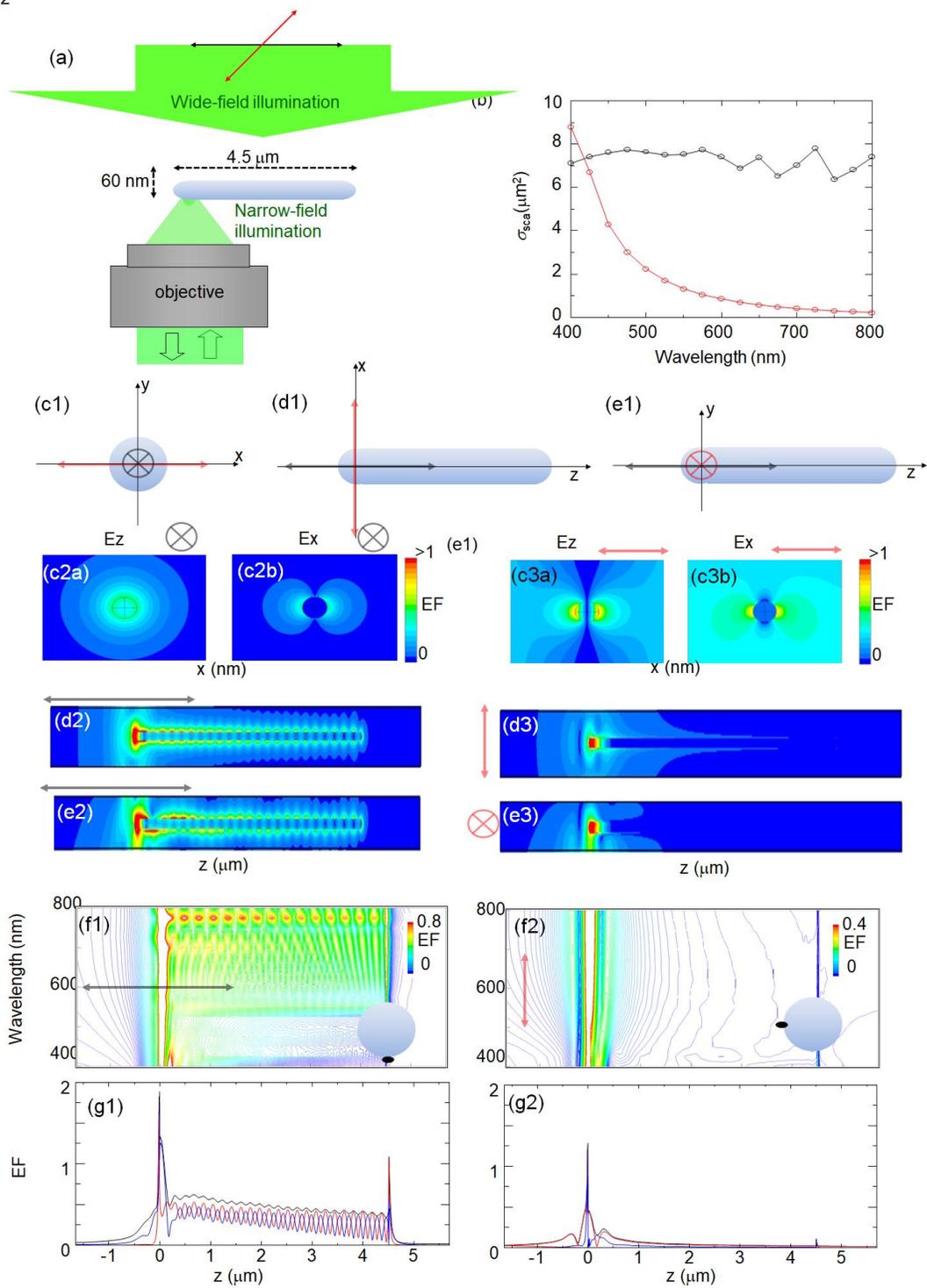

Fig. S3

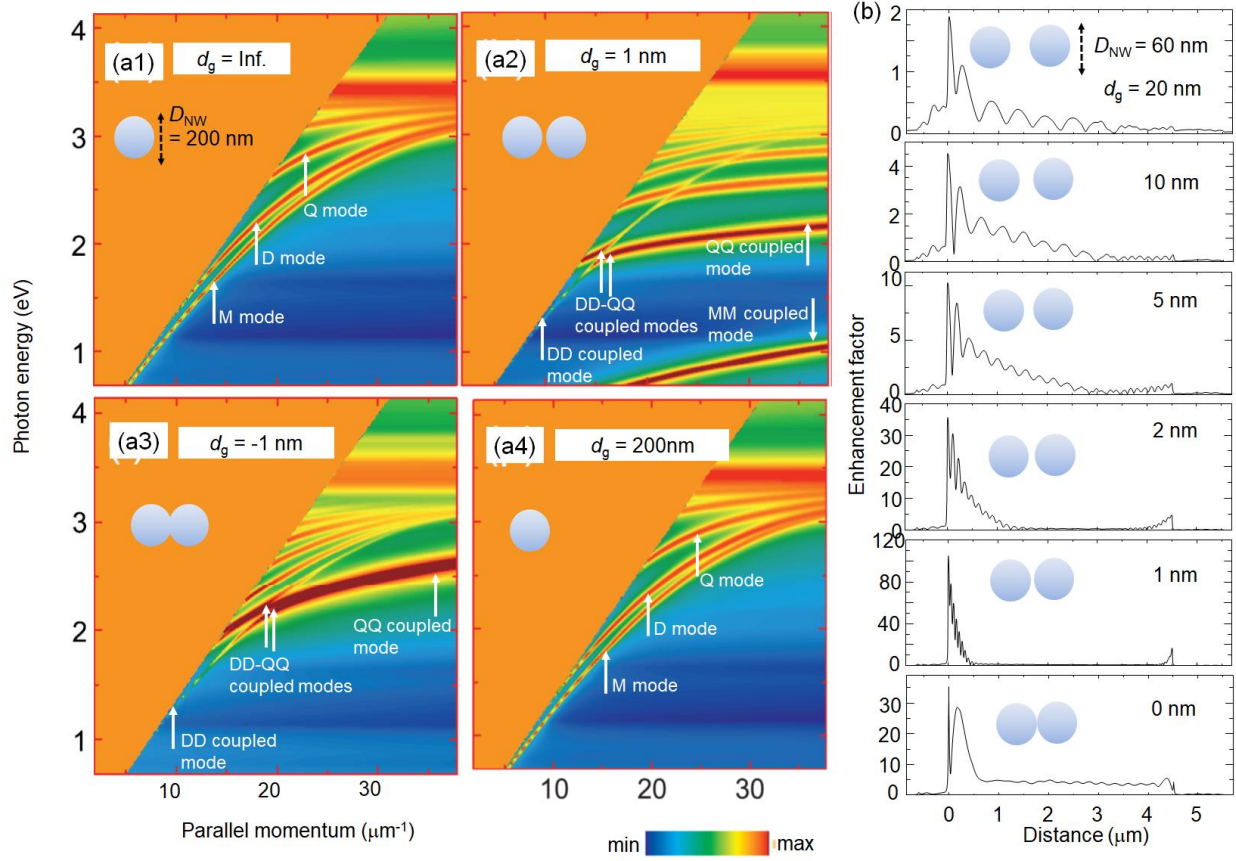